\documentclass{article}
\usepackage{spconf}

\usepackage{url}
\usepackage{amsmath,amssymb,amsfonts}
\usepackage{algorithmic}
\usepackage[linesnumbered,ruled,vlined]{algorithm2e}
\usepackage{graphicx}
\usepackage{textcomp}
\usepackage{colortbl, hhline}
\usepackage{xcolor}
\usepackage{multirow,tabularx}
\usepackage{makecell}
\usepackage{subcaption}
\usepackage[export]{adjustbox}
\usepackage[colorinlistoftodos]{todonotes}
\usepackage{verbatim}
\usepackage{hyperref}

\hypersetup{colorlinks=true,allcolors=blue}
\usepackage{color}

\definecolor{darkgreen}{rgb}{0., 0.4, 0.}
\definecolor{amber}{rgb}{1.0, 0.49, 0.0}
\definecolor{orange}{rgb}{1.0, 0.4, 0.0}
\definecolor{carmine}{rgb}{0.59, 0.0, 0.09}

\title{Tech Report: A Homogeneity-Based Multiscale Hyperspectral Image Representation for Sparse Spectral Unmixing}
\name{Luciano C. Ayres$^{\star}$ \qquad S\'ergio J. M. de Almeida$^{\star}$ \qquad Jos\'e C. M. Bermudez$^{\star \dagger}$ \qquad Ricardo A. Borsoi$^{\dagger \ddagger}$}
\address{$^{\star}$ Catholic University of Pelotas, Pelotas, RS, Brazil \\ 
         $^{\dagger}$ Federal University of Santa Catarina, Florian\'opolis, SC, Brazil \\ 
         $^{\ddagger}$ Universit\'e Côte d'Azur, CNRS, OCA, Nice, France \\
         {\normalsize E-mail: lucayress@gmail.com; sergio.almeida@ucpel.edu.br; j.bermudez@ieee.org; raborsoi@gmail.com.}}

\begin{document}
\ninept
\maketitle
\begin{abstract}
Several approaches have been proposed to solve the spectral \emph{unmixing} problem in hyperspectral image analysis. Among them the use of sparse regression techniques aims to characterize the abundances in pixels based on a large library of spectral signatures known \emph{a priori}. 
Recently, the integration of image spatial-contextual information significantly enhanced the performance of sparse unmixing.
In this work, we propose a computationally efficient multiscale representation method for hyperspectral data adapted to the unmixing problem. The proposed method is based on a hierarchical extension of the SLIC oversegmentation algorithm constructed using a robust homogeneity testing. The image is subdivided into a set of spectrally homogeneous regions formed by pixels with similar characteristics (\emph{superpixels}).
This representation is then used to provide prior spatial regularity information for the abundances of materials present in the scene, improving the conditioning of the unmixing problem. Simulation results illustrate that the method is capable of estimating abundances with high quality and low computational cost, especially in noisy scenarios.
\end{abstract}
\section{Introduction}

Hyperspectral imaging has been effectively employed in many remote sensing applications \cite{schowengerdt2006remote}. Recent developments extended this technology to several areas such as military surveillance and reconnaissance \cite{Shimoni2019}, food quality control \cite{liu2017hyperspectral} and medicine \cite{halicek2019cancer}. Due to the limited spatial resolution of hyperspectral sensors, the measured reflectance spectrum of a single pixel can be composed of contributions from different materials \cite{Bioucas-Dias2012}. \emph{Spectral unmixing} is the process by which the spectrum of a mixed pixel from a hyperspectral image (HI) is decomposed into a collection of spectral signatures of pure materials (\emph{endmembers}) and a set of corresponding fractional \emph{abundances}, the latter indicating what proportion of that pixel corresponds to each endmember \cite{Keshava2002}. 
Most solutions to the unmixing problem are based on a linear mixing model (LMM) \cite{Keshava2002, Dobigeon2014}, which considers each reflectance vector (HI pixel) to be formed by the linear combination of a finite number of endmembers, weighted by their fractional abundances. 

The LMM leads to fast and reliable unmixing results when the endmembers are accurately estimated~\cite{Bioucas-Dias2012}.
However, most algorithms that extract the endmembers directly from a given scene rely on the presence of pure pixels, or on the data not being heavily mixed~\cite{ma2013signal}. To circumvent this limitation, \emph{sparse regression-based} linear unmixing assumes that the reflectances of the observed pixels in an HI can be expressed as linear combinations of a small number of endmember signatures contained in a large spectral library known \emph{a priori} \cite{Iordache2011}. The unmixing problem consists of finding the subset of signatures in the library and their abundances to best represent each pixel. Employing spectral libraries avoids the need for estimating the number of endmembers and their spectral signatures. However, using large libraries makes the unmixing problem ill-posed, which in turn makes the solution highly sensitive to noise \cite{Iordache2010}.
Traditional sparse unmixing processes the spectrum of each observed image pixel independently, thus disregarding the spatial arrangement and the correlation between neighboring pixels \cite{Iordache2010, Iordache2014, Wang2016, Zheng2016}.
Integrating spatial-contextual information through regularizers can significantly enhance the  performance of sparse unmixing \cite{Iordache2012, Wang2017a, Zhang2018}. However, these techniques usually lead to computationally expensive algorithms.

Recently, a fast sparse unmixing algorithm (called MUA) was proposed to efficiently introduce spatial context in the unmixing problem \cite{Borsoi2019}. Using image segmentation and superpixels techniques \cite{MacQueenJamesandothers1967, Veganzones2014, achanta2012slic}, the problem is divided into two spatial domains or scales: one with the original image and another with its coarse representation formed by the average of the pixels in each superpixel. Spectral unmixing is first performed in the coarse domain, leading to initial abundance estimates, which are then used to regularize the optimization problem to be solved in the original scale. Using the Simple Linear Iterative Clustering (SLIC) \cite{achanta2012slic} oversegmentation algorithm, MUA proved to be capable of estimating abundances with the quality comparable (or even better in noisy scenarios) to the state-of-the-art S$^2$WSU \cite{Zhang2018}, with significantly lower execution time. Nevertheless, MUA performance relies strongly on two characteristics of the oversegmentation results to achieve meaningful performance improvements: the superpixels must group a large number of pixels, and the pixels within a superpixel must be spectrally homogeneous, except for the influence of noise. However, traditional superpixel (or image segmentation) algorithms \cite{MacQueenJamesandothers1967,Veganzones2014,achanta2012slic,Beucher1993,Shi2000,Felzenszwalb2004,Levinshtein2009} are not designed to optimize these criteria. Thus, MUA revealed a greater sensitivity to the image content when applied to images with irregular spatial compositions. This shows the need for developing multiscale representations that address the particular needs of the sparse unmixing problem.

This paper introduces the Homogeneity-based Multiscale sparse Unmixing Algorithm (HMUA) which leads to a multiscale representation of an HI which improves the efficiency of sparse unmixing. The proposed method is based on the SLIC oversegmentation algorithm and on a robust superpixel homogeneity test. We propose a superpixel homogeneity assessment based on the Euclidean distance between the median vector of a superpixel and each of its pixels. The objective is to reduce the effect of possible outliers in the test result. Those superpixels classified as non-homogeneous are subdivided  into smaller regions by successive rounds of oversegmentation. The final oversegmentation is then used in the first abundance estimation step (coarse domain) of MUA. This method better characterizes the spatial arrangement of the abundances in an HI with content distributed in regions of distinct sizes and shapes, allowing the improvement of the sparse unmixing performance of the original MUA for HIs with irregular spatial compositions.

The paper is organized as follows. Section \ref{sec:MUA} briefly reviews the MUA formulation for the sparse unmixing problem. Section \ref{sec:proposed_method} introduces the design of the proposed multiscale representation of the HI. Experimental results and discussion are presented in Section~\ref{sec:results} and conclusions in Section \ref{sec:conclusions}.

\section{Multiscale sparse Unmixing} \label{sec:MUA}

Consider the LMM \cite{Bioucas-Dias2012} of an observed hyperspectral image $\mathbf{Y} \in \mathbb{R}^{L \times N}$ with $L$ bands and $N$ pixels as $\mathbf{Y} = \mathbf{AX}+\mathbf{N}$, where abundances $\mathbf{X} \in \mathbb{R}^{P \times N}$ are subject to nonnegativity  constraints. Matrix $\mathbf{A} \in \mathbb{R}^{L \times P}$ denotes a spectral library containing $P$ endmember signatures and $\mathbf{N} \in \mathbb{R}^{L \times N}$ represents the modeling errors and additive noise. The multiscale decomposition process is defined as the spatial transformation promoted by the operator $\mathbf{W} \in \mathbb{R}^{N \times K}$, $K<N$, which is applied to both the HI and the abundances as
\begin{equation} \label{eq:sparse_Yc}
\mathbf{Y}_{\!\mathcal{C}} = \mathbf{YW}, \qquad \mathbf{X}_\mathcal{C} = \mathbf{XW},    
\end{equation}
where $\mathbf{Y}_{\!\mathcal{C}} \in \mathbb{R}^{L \times K}$, $\mathbf{X}_\mathcal{C} \in \mathbb{R}^{P \times K}$, and the $\mathcal{C}$ subscript refers to the new approximate (coarse) domain of the image. The relation $K<N$ means that there are fewer superpixels in $\mathbf{Y}_{\!\mathcal{C}}$ than pixels in $\mathbf{Y}$. Columns of $\mathbf{Y}_{\!\mathcal{C}}$ (resp. $\mathbf{X}_\mathcal{C}$) are the average of all pixels (resp. abundances) in each superpixel. Applying the operator $\mathbf{W}$ according to (\ref{eq:sparse_Yc}), the sparse unmixing problem in the approximated domain is written as
\begin{equation}\label{eq:opt_coarse_Xc}
\hat{\mathbf{X}}_\mathcal{C} = \underset{\mathbf{X}_{\mathcal{C}} \geq 0}{\operatorname{arg\ min}}\ \frac{1}{2}\|\mathbf{Y}_{\!\mathcal{C}}-\mathbf{A}\mathbf{X}_\mathcal{C}\|^2_F + \lambda_\mathcal{C}\|\mathbf{X}_\mathcal{C}\|_{1,1}.
\end{equation}
To use the coarse abundance estimate $\hat{\mathbf{X}}_\mathcal{C}$ in the original unmixing problem, it is necessary to map it back to the original spatial scale, denoted by $\mathcal{D}$, as $\hat{\mathbf{{X}}}_\mathcal{D} = \hat{\mathbf{{X}}}_\mathcal{C}\mathbf{W}^* \in \mathbb{R}^{P \times N}$. Operation $\hat{\mathbf{{X}}}_\mathcal{C}\mathbf{W}^*$ applies the average abundance value of each superpixel in $\hat{\mathbf{X}}_\mathcal{C}$ to every pixel in the corresponding superpixel region
in the original domain. Then, the $\hat{\mathbf{{X}}}_\mathcal{D}$ low-resolution abundance matrix is used to regularize the unmixing problem on the original scale:
\begin{equation}\label{eq:opt_original}
\hat{\mathbf{X}} = \underset{\mathbf{X} \geq 0}{\operatorname{min}}\ \frac{1}{2}\|\mathbf{Y}-\mathbf{A}\mathbf{X}\|^2_F + \lambda\|\mathbf{X}\|_{1,1} + \frac{\beta}{2}\|\hat{\mathbf{{X}}}_\mathcal{D}-\hat{\mathbf{{X}}} \|^2_F.
\end{equation}
Optimization problems (\ref{eq:opt_coarse_Xc}) and (\ref{eq:opt_original}) are solved by adapting the \mbox{SUnSAL} algorithm \cite{Iordache2010,Eckstein1992} as described in detail in \cite{Borsoi2019}. A pseudocode is shown \mbox{in Algorithm~\ref{alg:MUA}.}

\begin{algorithm}[ht] 
	\SetAlgoLined
	\KwIn{matrix $\mathbf{Y}$, $\mathbf{A}$, $\mathbf{W}$ and parameters $\lambda_\mathcal{C}$, $\lambda$, $\beta$.}
	
	Calculate $\mathbf{Y}_{\!\mathcal{C}} = \mathbf{YW}$\;
	
	Find $\hat{\mathbf{X}}_C$ by solving (\ref{eq:opt_coarse_Xc}) using the ADMM \;
	Compute $\hat{\mathbf{{X}}}_\mathcal{D} = \hat{\mathbf{{X}}}_\mathcal{C}\mathbf{W}^*$\;
	Find $\hat{\mathbf{X}}$ by solving (\ref{eq:opt_original}) using the ADMM\;
	\Return{\rm{the estimated abundance matrix} $\hat{\mathbf{X}}$};
	\newline\footnotesize{OBS: ADMM:Alternating Direction Method of Multipliers~\cite{Eckstein1992}}
	\caption{\emph{Multiscale Sparse Unmixing}} \label{alg:MUA}
\end{algorithm}

\section{Proposed Hierarchical Oversegmentation} \label{sec:proposed_method}

When $\mathbf{W}$ is constructed from superpixel or image segmentation algorithms, it may not group the pixels into spectrally homogeneous regions. This may compromise MUA's performance. 
To address this problem, we propose a new multiscale segmentation of hyperspectral images specifically designed for the sparse unmixing problem. 
The content in an HI may be distributed in regions of irregular sizes and shapes.
Hence, using an oversegmentation algorithm with a single superpixel size may not be appropriate to characterize the spatial arrangement of the abundances. Large (small) patterns should be segmented using large (small) superpixels. For this, we propose to oversegment the HI progressively in multiple scales as follows. Start with large superpixels. After each oversegmentation step, assess the homogeneity of these regions using an appropriate metric. Regions classified as non-homogeneous are then subjected to new rounds of oversegmentation with progressively smaller superpixel sizes. This way it is possible to generate an adequate number of homogeneous superpixels that adapt to the different pattern shapes in the HI. Starting from the original HI $\mathbf{Y}$, we propose to progressively decompose its non-homogeneous superpixels (up to $R$ representation scales) until a prescribed level of homogeneity is achieved. Figure~\ref{fig:metodo_esquema} illustrates multiscale decomposition of an HI with the SLIC algorithm for $R=2$. The red dots with gray regions (blue dots with white regions) represent superpixels classified as non-homogeneous (homogeneous). Blue (red) lines indicate the decomposition of non-homogeneous superpixels in scale $r$ into homogeneous (non-homogeneous) superpixels in scale $r+1$. The hierarchical oversegmentation is detailed in the following.

The steps of the proposed HMUA for HI oversegmentation and spectral unmixing are as follows:

\begin{figure}[hbt]
	\centerline{\includegraphics[width=25em]{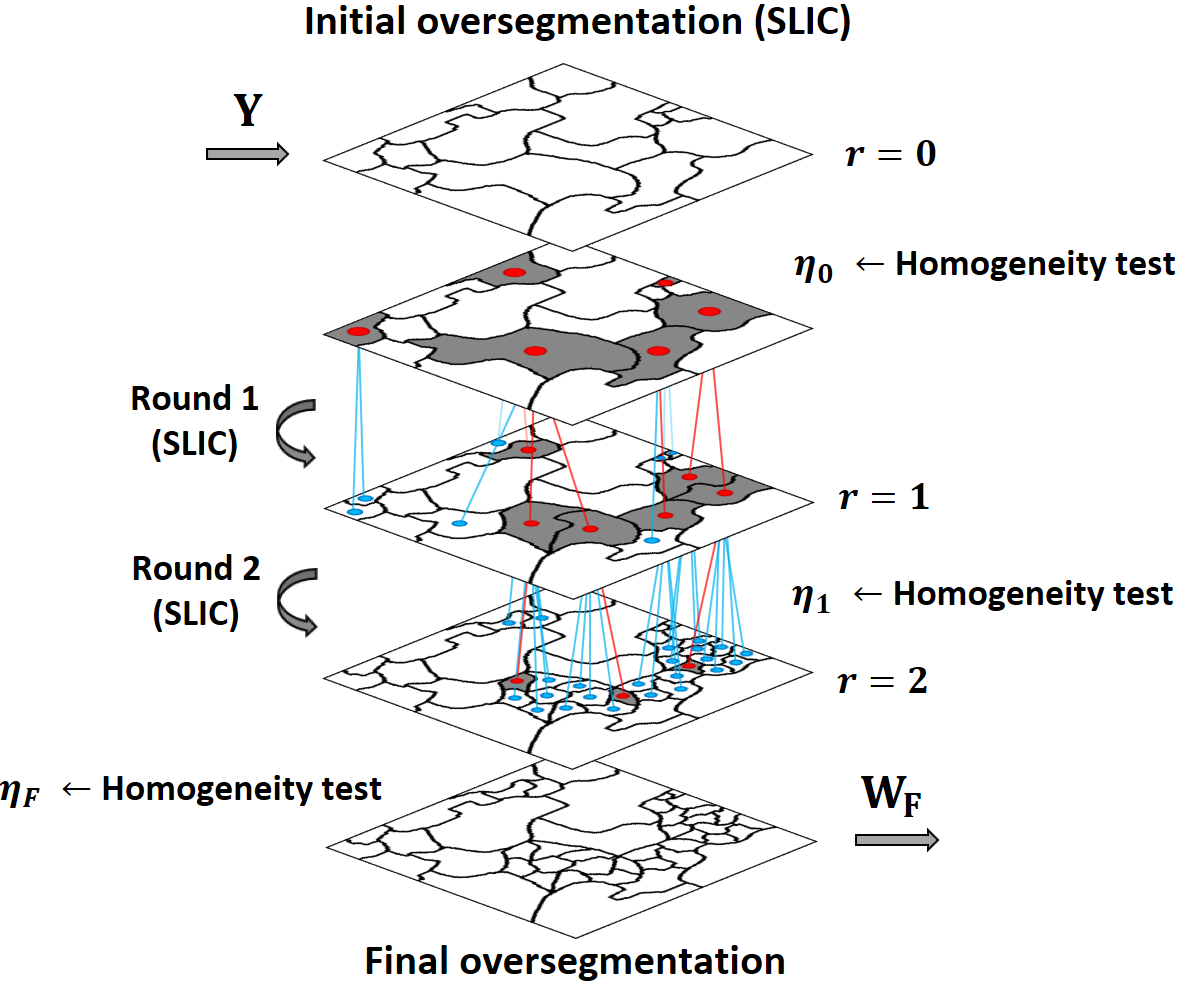}}
	\caption{Scheme of proposed method procedure with SLIC for $R=2$.
	}
	\label{fig:metodo_esquema}
\end{figure}

\textbf{Step 1} -- \emph{Initial image oversegmentation} 

The SLIC algorithm forms a segmentation map that subdivides the image into $K_0$ superpixels with average region size and compactness controlled by the parameters here defined as $\sigma_0$ and $\gamma$, respectively \cite{achanta2012slic}. 
Consider the matrix $\mathbf{S}_{r,k} \in \mathbb{R} ^{L\times|\mathcal{B}_{r,k}|}$ whose columns are the pixels contained in superpixel $k\in\{1, 2,\ldots,K_r\}$ in a scale of representation $r\in\{0, 1,\ldots,R\}$. The set $\mathcal{B}_{r,k}$ contains the indexes $I_n$, $n=1, 2,\ldots, |\mathcal{B}_{r,k}|$ ($|\cdot|$ denoting set cardinality) of each pixel in the $k^{\rm th}$ superpixel at the $r^{\rm th}$ representation scale. Then,
\begin{equation} \label{eq:sppx_S}
\mathbf{S}_{r,k} = \big[ \mathbf{y}_{I_1}, \mathbf{y}_{I_2},\ldots, \mathbf{y}_{I_{|\mathcal{B}_{r,k}|}} \big]
\end{equation}
where $\mathbf{y}_i$ is the $i^{\rm th}$ column of $\mathbf{Y}$ and $\{I_1, I_2,\ldots, I_{|\mathcal{B}_{r,k}|}\} \ = \mathcal{B}_{r,k}$. We also define as $\mathbf{W}_r$ the transformation promoted by the segmentation map $\mathbf{S}_{r,k}$ at the $r^{\rm th}$ representation level, for $r\in\{0,\ldots,R\}$.

\textbf{Step 2} -- \emph{Superpixels homogeneity test}

Let $\mathbf{m}_k$ be the median of the pixels in $\mathbf{S}_{r,k}$, at a given representation scale $r$:
\begin{equation} \label{eq:sppx_m}
\mathbf{m}_k = 
\begin{bmatrix}
m_1 \\
\vdots \\
m_L
\end{bmatrix} =
\begin{bmatrix}
\operatorname{med}([y_{1,I_1},\ldots,y_{1,I_{|\mathcal{B}_{r,k}|}}]) \\
\vdots \\
\operatorname{med}([y_{L,I_1},\ldots,y_{L,I_{|\mathcal{B}_{r,k}|}}])
\end{bmatrix},
\end{equation}
where $y_{\ell,i}$ is the $\ell^{\rm th}$ band of $\mathbf{y}_i$.
Define $\mathbf{d}_k$ as the vector of Euclidean distances between each pixel $\mathbf{y}_{I_n}$ and $\mathbf{m}_k$:
\begin{equation}\label{eq:sppx_dist_px}
\mathbf{d}_k=\big[d_1,d_2,\ldots,d_{|\mathcal{B}_{r,k}|} \big]^{\top}, \quad
d_n=\|\mathbf{m}_k - \mathbf{y}_{I_n}\|_{2}  \,.
\end{equation}
In measuring homogeneity, one must note that the influence of the noise $\mathbf{N}$ in $\mathbf{Y}$ means that the values in $\mathbf{d}_k$ are, on average, bounded away from zero even for perfectly homogeneous superpixels. Moreover, the presence of a small number of outliers can disproportionately bias the estimate. To address these problems, we first eliminate possible outliers by removing a percentage $\tau_{\text{outliers}}$ of the highest values from $\mathbf{d}_k$. This results in a reduced distance vector $\mathbf{d}_{k}' \in \mathbb{R}^{\lfloor D \rfloor}$ with $D=(1-\tau_{\text{outliers}})|\mathcal{B}_{r,k}|$, $\lfloor\cdot\rfloor$ being the floor function. The homogeneity measure $\delta_k$ is then defined as the deviation between the maximum distance $\max(\mathbf{d}_{k}')$ found in the $k^{\rm th}$ superpixel after removing the outliers, with respect to the average $\overline{\mathbf{d}_{k}'}$ of its distances:
\begin{equation}\label{eq:sppx_delta}
\delta_k=\frac{\max(\mathbf{d}_{k}')-\overline{\mathbf{d}_{k}'}}{\overline{\mathbf{d}_{k}'}}, \qquad \text{Homogeneous:  }\delta_k \leq \tau_{\text{homog}} ,
\end{equation}
Superpixels are classified as homogeneous if $\delta_k$ is below an acceptable threshold~$\tau_{\text{homog}}$. 
The percentage of homogeneous superpixels in the $r^{\rm th}$ representation scale with $K_r$ superpixels is given by $\eta_r = ({K_H}/K_r)\times 100\%$, where $K_H$ is the number of superpixels classified as homogeneous.

\textbf{Step 3} -- \emph{Subdivision of non-homogeneous superpixels}

Regions classified as non-homogeneous in Step~2 are submitted to an additional oversegmentation step with a smaller average region size parameter $\sigma_r<\sigma_{r-1}$, $\forall r$. This process is repeated for $r=1,\ldots,R$, or until $\eta_r = 100\%$ is reached. This generates a sequence of segmentation maps $\mathbf{S}_{1,k},\mathbf{S}_{2,k},\ldots,\mathbf{S}_{R,k}$ with an increasing level of spatial definition and superpixel homogeneity. The final transformation operator related to $\mathbf{S}_{r,k}$ is denoted $\mathbf{W}_F$.

\textbf{Step 4} -- \emph{Sparse unmixing}

The hierarchical oversegmentation yields the spatial transformation operator $\mathbf{W}_F$, which can then be used with the MUA strategy to estimate the abundance matrix $\hat{\mathbf{X}}$. A pseudocode for the HMUA sparse unmixing method is presented in Algorithm \ref{alg:metodo}.

\begin{algorithm}[h!] 
	\SetAlgoLined
	\KwIn{hyperspectral image $\mathbf{Y}$, spectral library $\mathbf{A}$, parameters $\gamma$, $\sigma_0$, $\sigma_1$, $\ldots$, $\sigma_R$, $\tau_{\text{outliers}}$, $\tau_{\text{homog}}$, $\lambda_\mathcal{C}$, $\lambda$,~$\beta$.}
	$\mathbf{W}_0, \mathbf{S}_{0,k} \leftarrow$ initial oversegmentation of $\mathbf{Y}$\;
	$\mathbf{W}_F \leftarrow \mathbf{W}_0$\;
	$\eta_0 \leftarrow$ homogeneity test of $\mathbf{S}_{0,k}$\;
	\For{$r=1$ \KwTo $R$}{
		\If{$\eta_{r-1} < 100\%$}{
			$\mathbf{W}_r, \mathbf{S}_{r,k} \leftarrow$ oversegmentation of non-homogeneous superpixels of $\mathbf{S}_{r-1,k}$ with $\sigma_r < \sigma_{r-1}$\;
			$\eta_r \leftarrow$ homogeneity test of $\mathbf{S}_{r,k}$\;
			$\mathbf{W}_F \leftarrow \mathbf{W}_r$\;
		}
	}
	$\hat{\mathbf{X}} \leftarrow$ Algorithm \ref{alg:MUA} (MUA) with $\mathbf{W} \equiv \mathbf{W}_F$\;    
	\Return{\rm{the estimated abundance matrix} $\hat{\mathbf{X}}$};
	\caption{\textit{Homogeneity-based Multiscale sparse Unmixing}} \label{alg:metodo}
\end{algorithm}

\vspace{-1ex}
\section{Experimental Results and Discussion} \label{sec:results}

We compared the proposed HMUA with $\text{MUA}_\text{SLIC}$ \cite{Borsoi2019} and S$^2$WSU \cite{Zhang2018} in terms of abundance estimation quality and computational complexity. The choice of these algorithms is justified by the fact that they have already proven a better performance over others of the same class, e.g., SUnSAL \cite{Iordache2010}, SUnSAL-TV \cite{Iordache2012}, DRSU \cite{Wang2016} and DRSU-TV \cite{Wang2017a}. We used synthetic data to facilitate objective evaluations. Results with real data corroborated the conclusions and are shown in \cite{supp} due to space limitations.

\vspace{-2ex}
\subsection{Configuration and datasets}

\begin{figure}[htb]
	\centering
	\begin{subfigure}[b]{0.13\textwidth}
		\centerline{\includegraphics[width=6em]{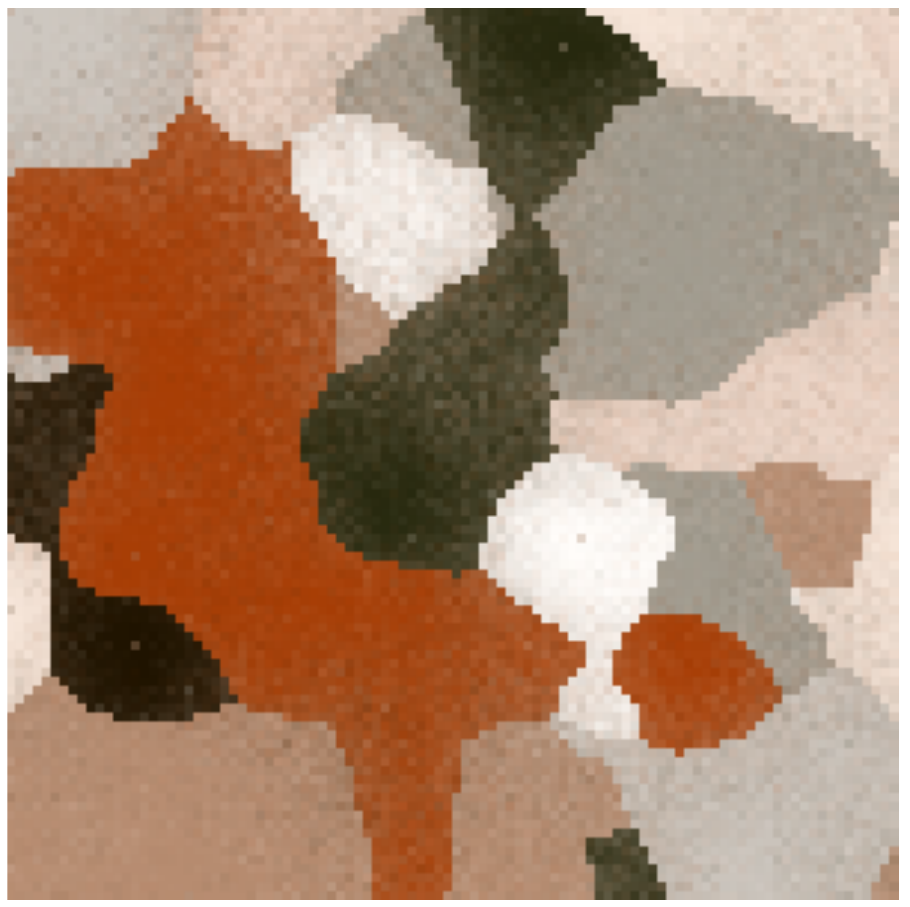}}
		\caption{DC1}
		\label{fig:DC1_RGB}
	\end{subfigure}
	\begin{subfigure}[b]{0.13\textwidth}
		\centerline{\includegraphics[width=6em]{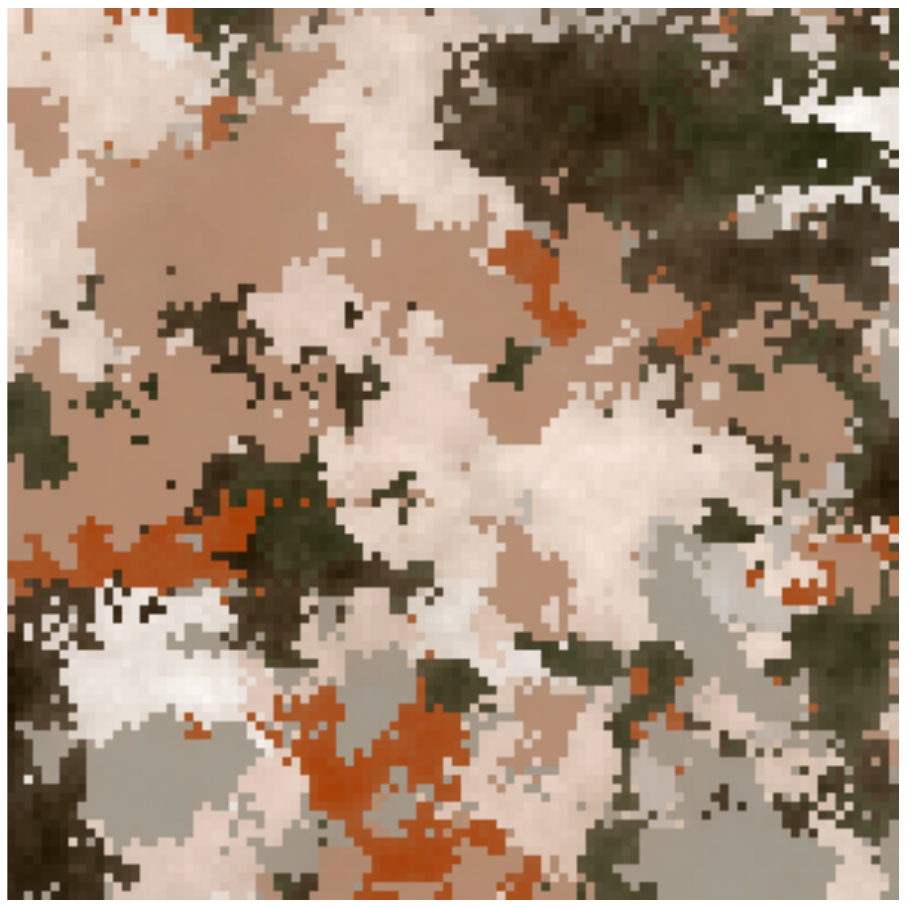}}
		\caption{DC2}
		\label{fig:DC2_RGB}
	\end{subfigure}
	\begin{subfigure}[b]{0.13\textwidth}
		\centerline{\includegraphics[width=6em]{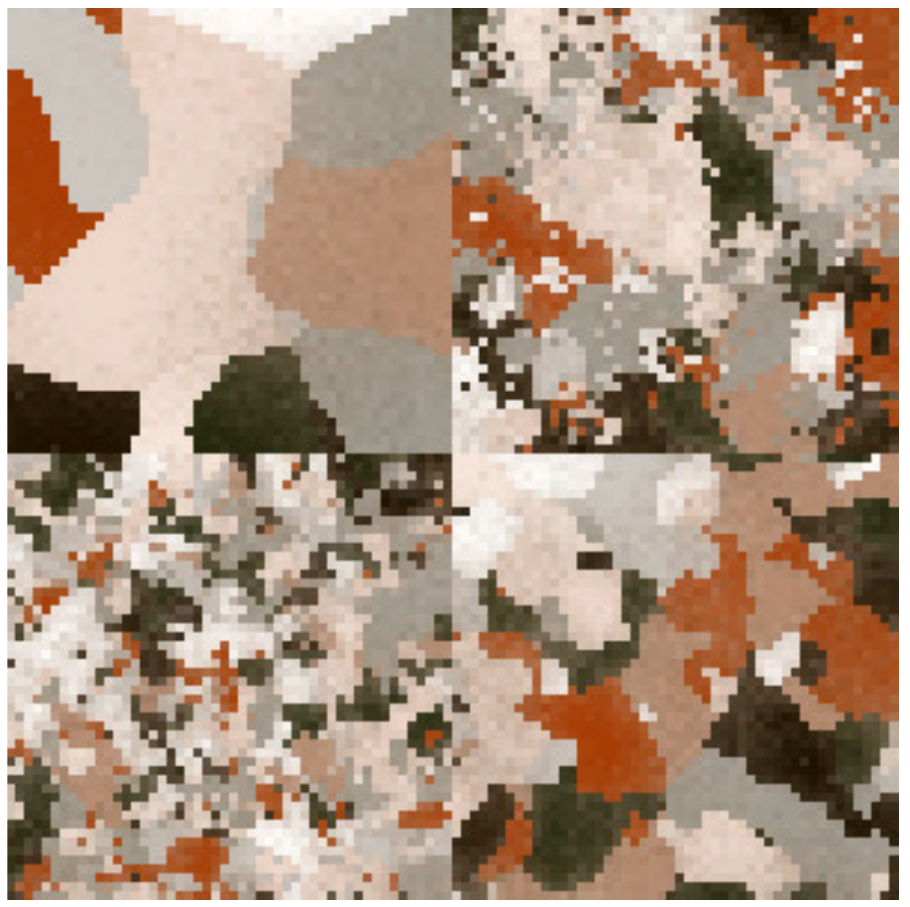}}
		\caption{DC3}
		\label{fig:DC3_RGB}
	\end{subfigure}
	\vspace{-0.2cm}
	\caption{RGB representation of the synthetic HIs.}
	\label{fig:DCs_synth}
\end{figure}

For the results to be compared to a known reference (ground-truth), three synthetic HIs (DC1, DC2 and DC3, represented in Figure \ref{fig:DCs_synth}) with $100 \times 100$ pixels were generated using nine endmembers selected from a library $\mathbf{A} \in \mathbb{R}^{224 \times 240}$ composed of a subset of 240 materials signatures from the USGS \emph{splib06}\footnote{Available at www.usgs.gov/labs/spec-lab/capabilities/spectral-library.} library. To test the techniques in different scenarios, the data cubes have varied spatially correlated abundance distributions generated by the \emph{Hyperspectral Imagery Synthesis}\footnote{Available at www.ehu.eus/ccwintco.} tool: DC1 -- medium and large uniform areas; DC2 -- regions of irregular size and contours and DC3 -- a composite of four $50 \times 50$ pixels images of varied arrangements. 
Finally, white Gaussian noise was added to the generated images to obtain signal-to-noise ratios (SNR) of 20 and 30 dB. {The low SNR condition is justified especially considering cases of low-cost sensors that may not have such a high SNR.}
Transformation $\mathbf{W}$ was performed by SLIC oversegmentation implemented in the \emph{VLFeat} toolbox \cite{Vedaldi2008}, which allows its use for multichannel data.
SLIC works with two parameters: $\sigma$, which sets the average size $\sqrt{N/K}$ of the superpixels, and $\gamma$, which determines the weight of the spatial contribution in the pixel similarity metric \cite{achanta2012slic}.
As a quantitative criterion of the unmixing performance, we use the signal-to-reconstruction error \cite{Iordache2012}, $\text{SRE (dB)} = 10\log_{10}(\mathbb{E}\|\mathbf{X}\|^2_F/\mathbb{E}\|\mathbf{X}-\hat{\mathbf{X}}\|^2_F$), which assesses the estimation of the abundances.
All results are based on optimal parameter values obtained for each HI. To find them, a grid search was performed in the following ranges: Multiscale representation -- rounds $R=3$, regularizer $\gamma \in \{0.00025,0.00125,\ldots,0.1\}$, superpixels size $\sigma_0, \sigma_1, \sigma_2, \sigma_3, \in \{5,6,\ldots,14\}$, considering  $\sigma_0 > \sigma_1 > \sigma_2 > \sigma_3$ and thresholds $\tau_{\text{outliers}} \in \{10\%,20\%,30\%\}$, $\tau_{\text{homog}} \in \{10\%,20\%,\ldots,60\%\}$; Sparse unmixing -- regularizers $\lambda_{\mathcal{C}}$, $\lambda$ and $\lambda_{\text{S²WSU}}$ were varied according to the values $1,3,5,7,9 \times 10^i$, for $i \in \{-3,-2,-1,0\}$ and $\beta$ in $1,3,5 \times 10^j$, for $j \in \{-1,0,1,2\}$. Algorithms were executed in \emph{MATLAB$^{\text{\tiny TM}}$}, on a computer equipped with an \emph{Intel Core i7 3537U @ 2.00GHz} processor and 8GB RAM.

\vspace{-0.75cm}
\subsection{Results and discussion}
\begin{figure}[h!]
	\centering
	\begin{subfigure}[b]{0.2\textwidth}
		\centerline{\includegraphics[width=9em]{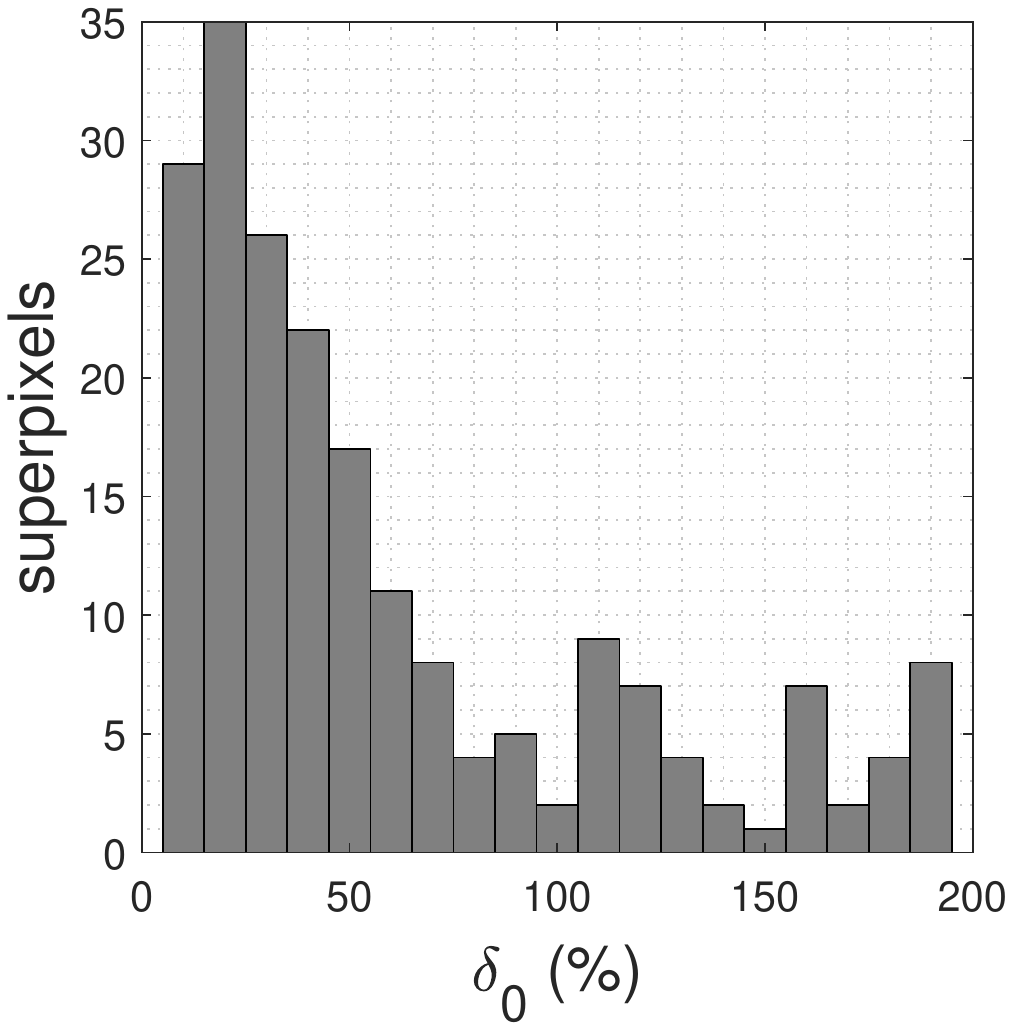}}
		\label{fig:DC3_30dB_ratio_inicial}
	\end{subfigure}
	\begin{subfigure}[b]{0.2\textwidth}
		\centerline{\includegraphics[width=9.2em]{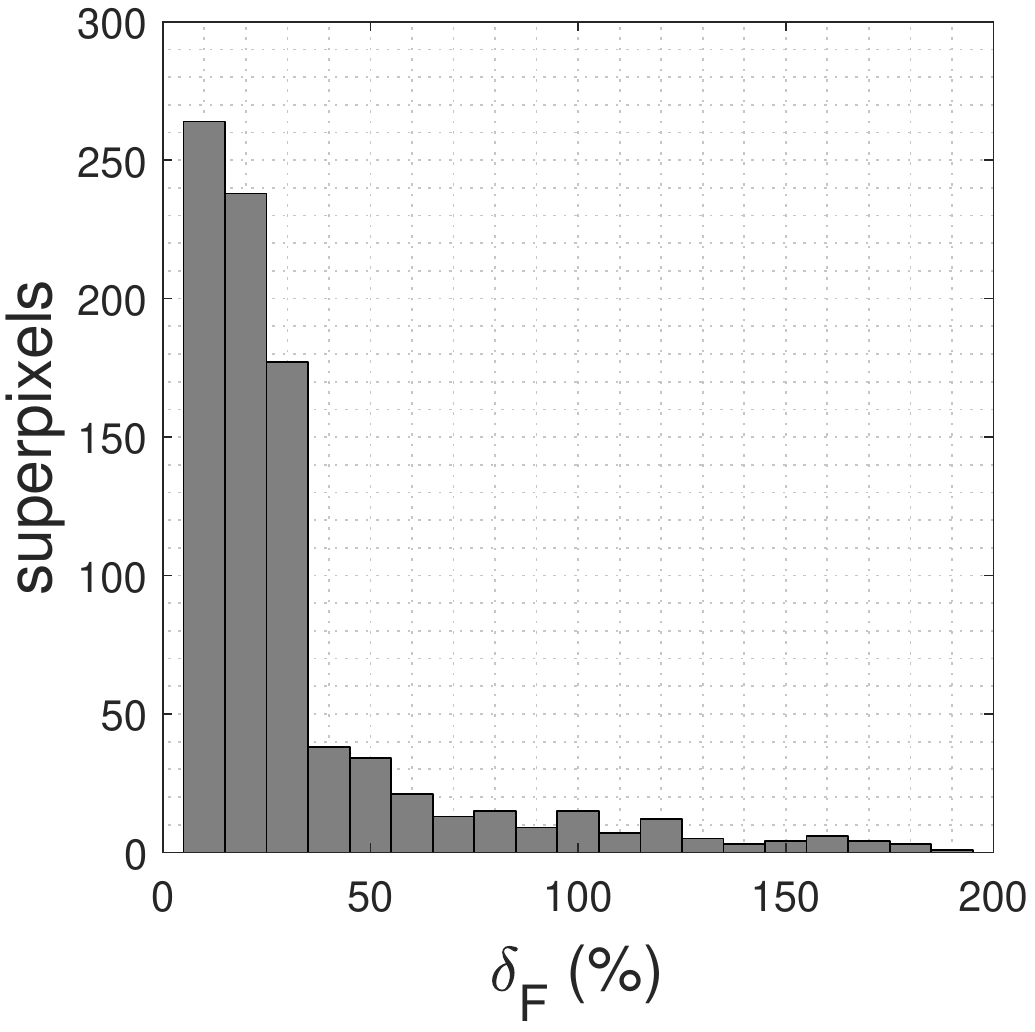}}
		\label{fig:DC3_30dB_ratio_final}
	\end{subfigure}
	\vspace{-0.2cm}
	\caption{Comparison between initial (left) and final (right) $\delta$ deviation values of superpixels in DC3, SNR 30 dB.}
	\label{fig:DC3_homog_30dB_results}
\end{figure}

Despite the number of parameters to set in HMUA, through sensitivity and statistical performance analyses, we concluded that there is a high chance of obtaining good results by choosing values within a reasonable range. These analyses and the optimal parameter values found are shown in \cite{supp} due to space limitations. In all cases removing only $10\%$ of the highest values from $\mathbf{d}_k$ in \eqref{eq:sppx_dist_px} ($\tau_{\text{outliers}}=0.1$) was sufficient to avoid outliers in homogeneity tests. Also, it can be seen from Table~\ref{tab:comparacao_algoritmos_superpixels} that in each scene there was an increase in the percentage of homogeneous superpixels from the initial ($\eta_0$) towards the final ($\eta_F$) oversegmentation round, particularly for DC2 and DC3. The histograms in Figure~\ref{fig:DC3_homog_30dB_results} illustrate this evolution in terms of the values of $\delta_k$.
Moreover, only three rounds of oversegmentation ($R=3$) were sufficient to obtain a high $\eta_F$ index and an adequate unmixing result, without compromising the low computational cost of the method. Longer execution times are related to a low value of $\eta_0$, as it requires a greater number of superpixel homogeneity evaluations, subdivisions and unmixing.

\begin{figure}[t!]
	\centerline{\includegraphics[width=\linewidth]{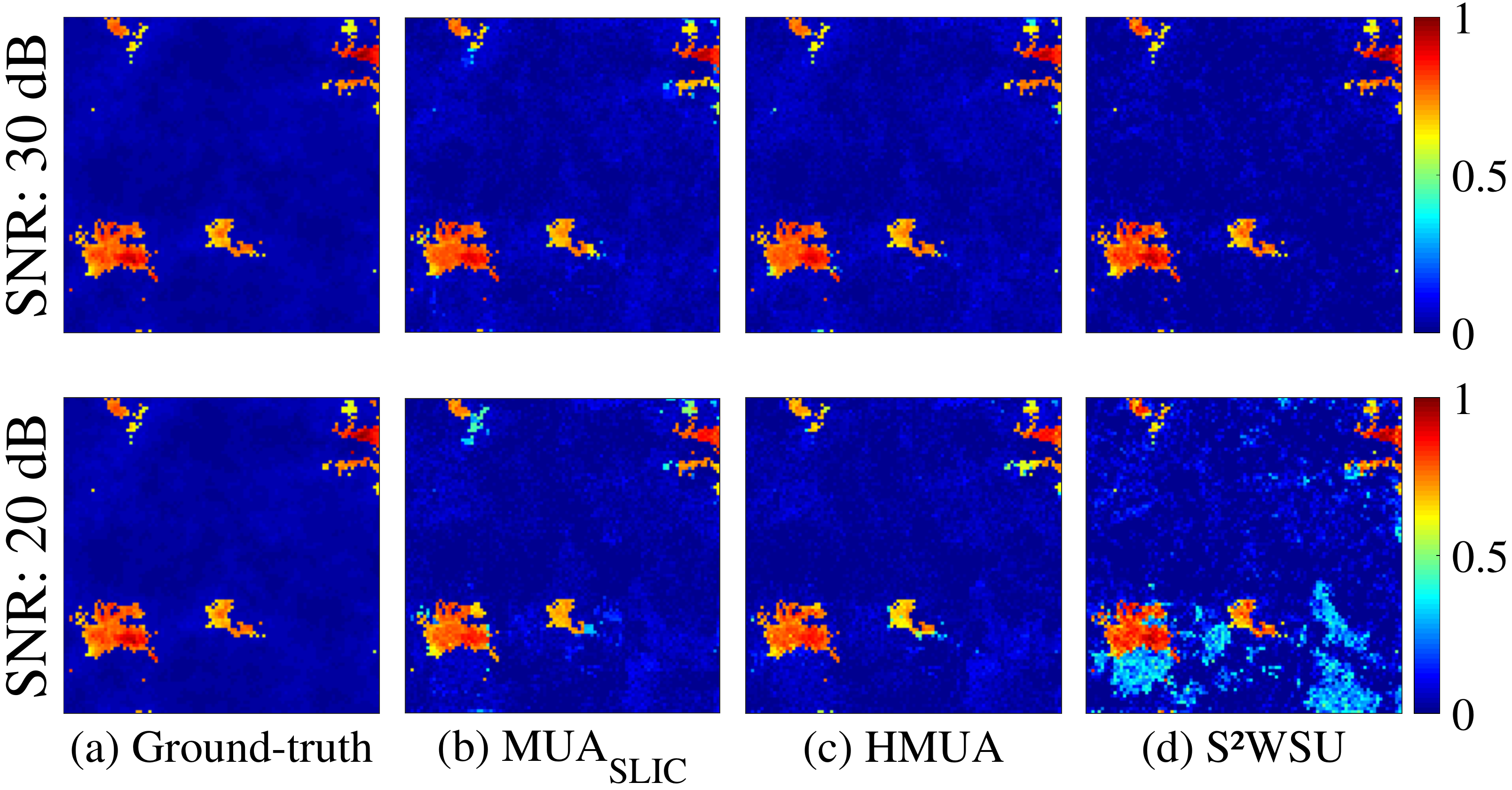}}
	\vspace{-0.35cm}
	\caption{Abundance estimation results for endmember 3 of DC2.
	}
	\label{fig:DC2_comaparacao}
\end{figure}

Table~\ref{tab:comparacao_algoritmos_superpixels} shows the number of superpixels used with the original $\text{MUA}_\text{SLIC}$ and with the HMUA. The proposed method was capable of reducing the necessary amount of superpixels in more uniform scenarios, such as that of DC1. Moreover, the final oversegmentation given by the proposed approach presented a considerably higher percentage of homogeneous superpixels in the images.
Table~\ref{tab:comparacao_algoritmos_SRE} shows the SRE performance of the algorithms compared to the optimal value. The HMUA obtained the best quantitative results in a noisy scenario (SNR 20 dB). A noticeable improvement was verified for DC3, with an approximate 5\% increase in SRE for both 20 and 30 dB SNR conditions when compared to $\text{MUA}_\text{SLIC}$. 
These results indicate that the HMUA tends to be more effective when the characteristics of the abundances spatial content varies across the scene. 
The execution time of the proposed HMUA, shown in Table~\ref{tab:comparacao_algoritmos_tempo}, was similar to that of $\text{MUA}_\text{SLIC}$, and significantly lower than that of S$^2$WSU, despite the additional steps of homogeneity assessment and extra oversegmentations.
Figure~\ref{fig:DC2_comaparacao} shows the true and reconstructed abundances maps of endmember 3 of DC2. 
The abundances estimated by S$^2$WSU are visually the closest to the ground truth for a high SNR (30 dB), while the results by the HMUA and $\text{MUA}_\text{SLIC}$ were more similar. However, for an SNR of 20 dB, the performance of S²WSU degrades sharply, and the results by the HMUA show a clear improvement over $\text{MUA}_\text{SLIC}$, which illustrates its effectiveness for noisier images.

\begin{table}[t!]
	\centering
	\caption{Number of generated superpixels.}
	\vspace{-0.2cm}
	\centering
	\begin{tabular}{c|c|c|c|c|c} 
		\hline
		\multirow{2}{*}{Data} & \multirow{2}{*}{SNR} & \multicolumn{2}{c|}{$\text{MUA}_\text{SLIC}$} & \multicolumn{2}{c}{HMUA}     \\
		\cline{3-6}
		& & \emph{superpixels} & $\eta\equiv\eta_0$ & \emph{superpixels} & $\eta\equiv\eta_F$ \\
		\hline
		\multirow{2}{*}{DC1} & 30 dB & 225 & 89\% & \textbf{218} & \textbf{99\%}  \\
		\cline{2-6}
		& 20 dB & 169 & 97\% & \textbf{105} & \textbf{98\%}  \\
		\hline
		\multirow{2}{*}{DC2} & 30 dB & \textbf{625} & 69\% & 1018 & \textbf{90\%} \\
		\cline{2-6}
		& 20 dB & \textbf{289} & 81\% & 444  & \textbf{94\%} \\
		\hline
		\multirow{2}{*}{DC3} & 30 dB & \textbf{624} & 55\% & 1566  & \textbf{84\%} \\
		\cline{2-6}
		& 20 dB & 625 & 80\% & \textbf{610}  & \textbf{84\%} \\
		\hline
	\end{tabular}
	\label{tab:comparacao_algoritmos_superpixels}
\end{table}

\begin{table}[t!]
	\centering
	\caption{SRE results.}
	\vspace{-0.35cm}
	\centering
	\begin{tabular}{c|c|c|c|c} 
		\hline
		Data & SNR & S$^2$WSU & $\text{MUA}_\text{SLIC}$ & HMUA    \\
		\hline
		\multirow{2}{*}{DC1} & 30 dB &  \textbf{21,668 dB} & 18,117 dB & 18,339 dB \\ 
		\cline{2-5}
		& 20 dB &   9,332 dB & 14,854 dB & \textbf{15,104 dB} \\  
		\hline
		\multirow{2}{*}{DC2} & 30 dB & \textbf{18,741 dB} & 11,737 dB & 11,780 dB \\
		\cline{2-5}
		& 20 dB &  5,689 dB &  8,416 dB &  \textbf{8,561 dB} \\
		\hline
		\multirow{2}{*}{DC3} & 30 dB & \textbf{19,798 dB} & 10,841 dB & 11,398 dB \\
		\cline{2-5}
		& 20 dB &  6,899 dB &  7,776 dB &  \textbf{8,185 dB} \\
		\hline
	\end{tabular}
	\label{tab:comparacao_algoritmos_SRE}
\end{table}

\begin{table}[t!]
	\centering
	\caption{Average execution times.}
	\vspace{-0.35cm}
	\centering
	\begin{tabular}{c|c|c|c|c} 
		\hline
		Data & SNR & S$^2$WSU & $\text{MUA}_\text{SLIC}$ & HMUA    \\
		\hline
		\multirow{2}{*}{DC1} & 30 dB & 239 s & \textbf{9  s} & 11 s \\
		\cline{2-5}
		& 20 dB & 235 s & \textbf{15 s} & 16 s \\
		\hline
		\multirow{2}{*}{DC2} & 30 dB & 233 s & \textbf{10 s} & 16 s \\
		\cline{2-5}
		& 20 dB & 232 s & \textbf{10 s} & 16 s \\
		\hline
		\multirow{2}{*}{DC3} & 30 dB & 233 s & \textbf{12 s} & 21 s \\
		\cline{2-5}
		& 20 dB & 232 s & \textbf{12 s} & 15 s \\
		\hline
	\end{tabular}
	\label{tab:comparacao_algoritmos_tempo}
\end{table}

\vspace{-0.3cm}
\section{Conclusions} \label{sec:conclusions}

In this work we proposed a new hierarchical multiscale representation method for sparse unmixing of HIs. Using the SLIC oversegmentation algorithm and a novel robust homogeneity testing methodology, we iteratively divide an HI into irregular superpixels with improved spectral homogeneity, which better reflect the spatial information of the abundances in mixed pixels of the HI. The final superpixel decomposition is then used to introduce spatial information into sparse unmixing by means of a multiscale algorithm. Experimental results with hyperspectral data of distinct spatial compositions showed that the HMUA outperforms state-of-the-art algorithms in noisy scenarios while still maintaining a very low computational complexity.

\vfill\pagebreak

\bibliographystyle{IEEEtran}
\bibliography{library}

\vfill\pagebreak

\onecolumn
\title{\textbf{Supplemental Material}}

The following material supplements the paper
\vspace{0.5cm}
\newline
[A] L. C. Ayres, S. J. M. de Almeida, J. C. M. Bermudez and R. A. Borsoi, "A Homogeneity-based Multiscale Hyperspectral Image Representation for Sparse Spectral Unmixing". \textit{2021 IEEE International Conference on Acoustics, Speech and Signal Processing}.

\section{Sensitivity analysis}

\begin{figure}[h!]
	\centering
	\begin{subfigure}[b]{0.29\textwidth}
		\centerline{\includegraphics[width=10em]{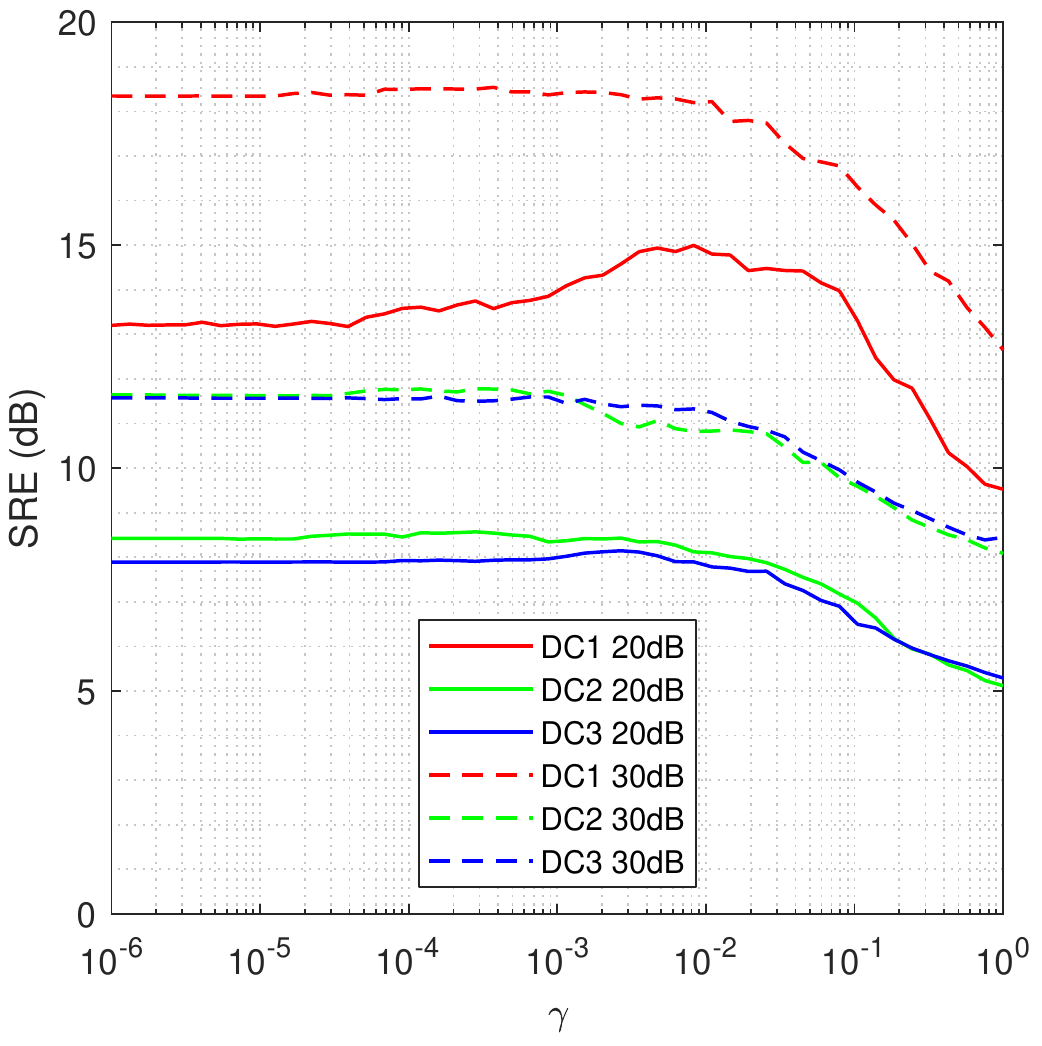}}
		\label{fig:DCs_sensitivity_gamma}
	\end{subfigure}
	\begin{subfigure}[b]{0.29\textwidth}
		\centerline{\includegraphics[width=10em]{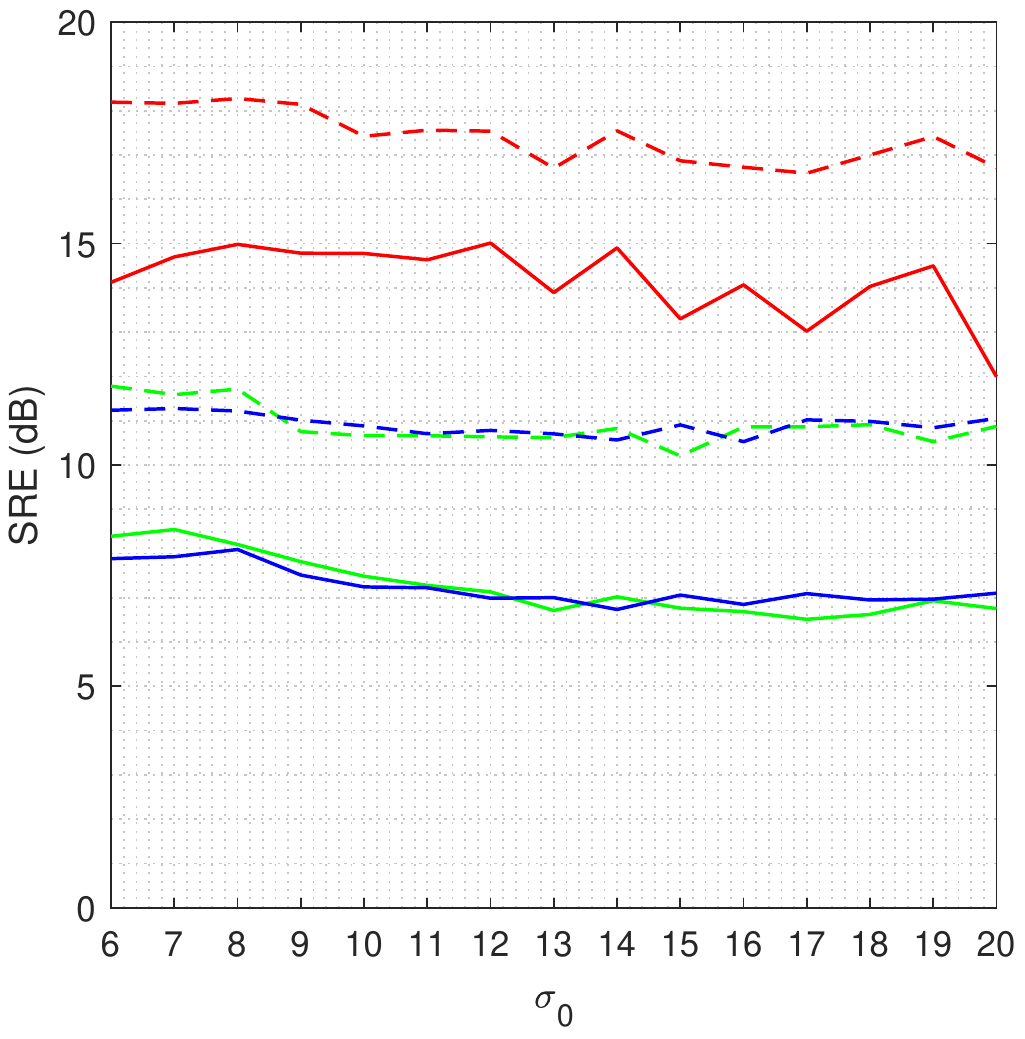}}
		\label{fig:DCs_sensitivity_sigma_SRE}
	\end{subfigure}

	\begin{subfigure}[b]{0.29\textwidth}
		\centerline{\includegraphics[width=10em]{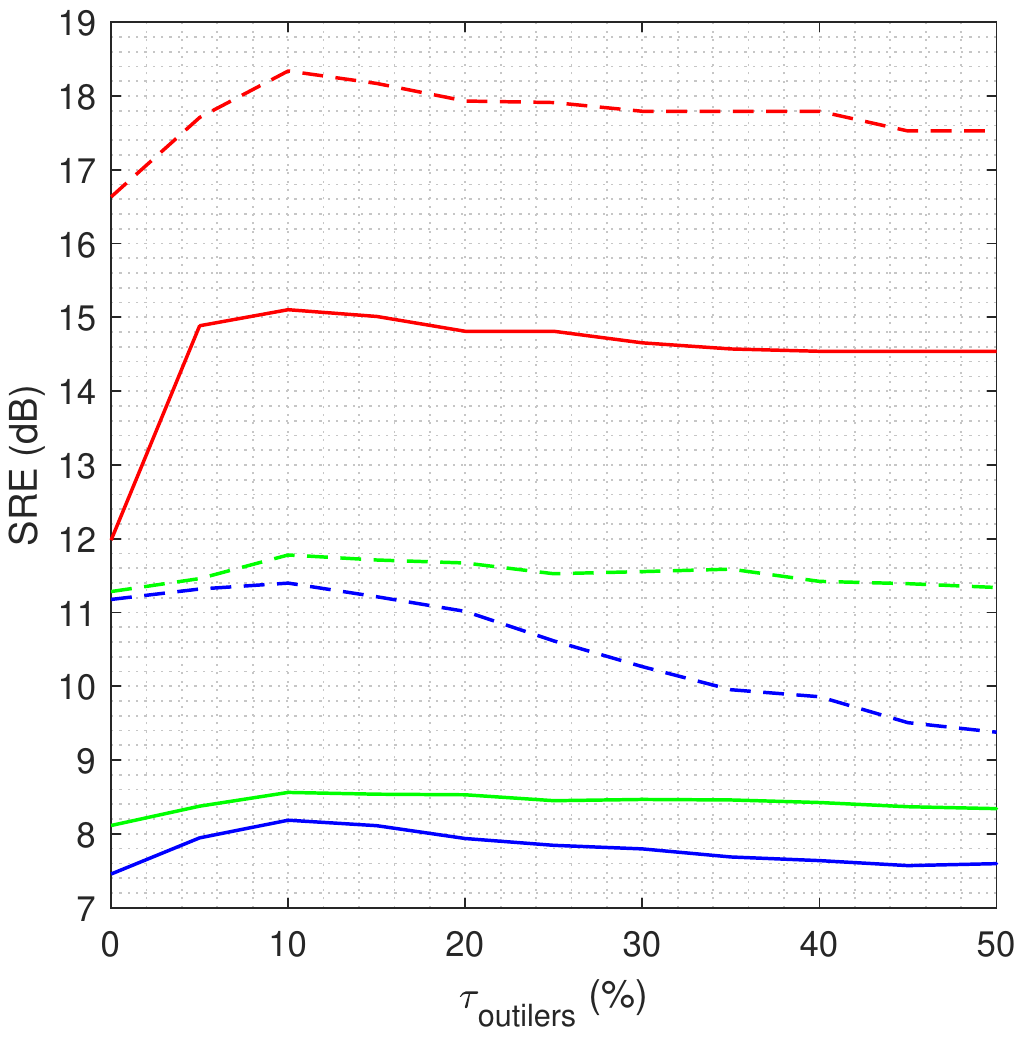}}
		\label{fig:DCs_sensitivity_tau_outliers}
	\end{subfigure}
	\begin{subfigure}[b]{0.29\textwidth}
		\centerline{\includegraphics[width=10em]{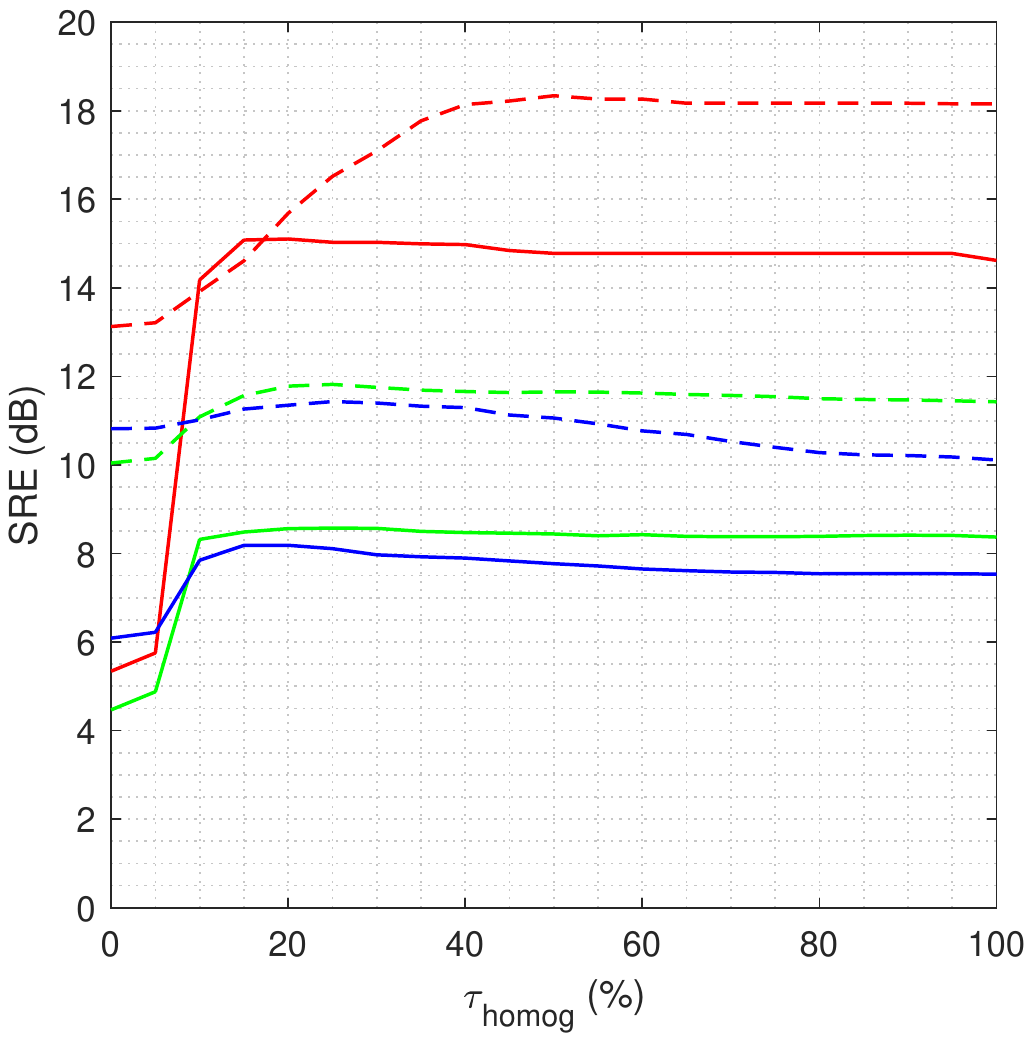}}
		\label{fig:DCs_sensitivity_tau_homog}
	\end{subfigure}
	
	\begin{subfigure}[b]{0.29\textwidth}
		\centerline{\includegraphics[width=10em]{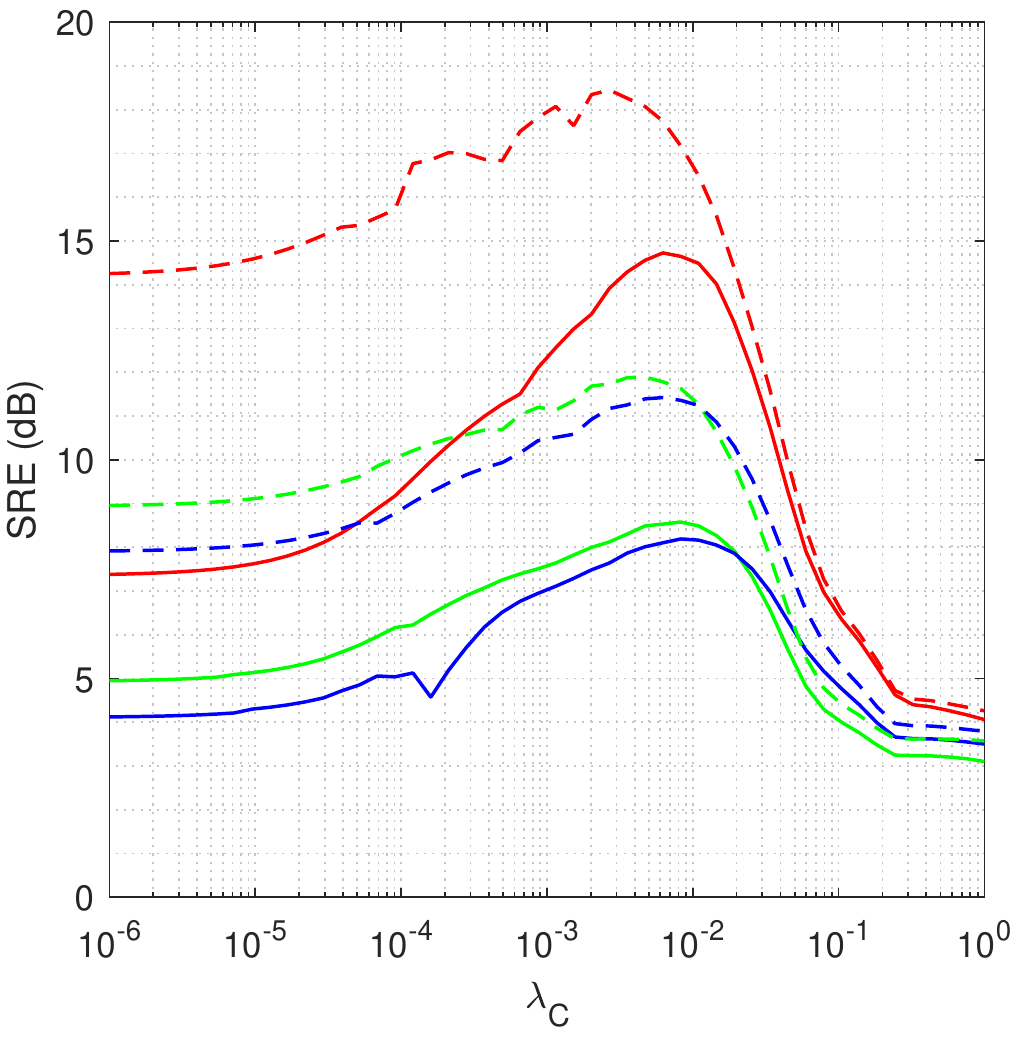}}
		\label{fig:DCs_sensitivity_lambda_c}
	\end{subfigure}
	\begin{subfigure}[b]{0.29\textwidth}
		\centerline{\includegraphics[width=10em]{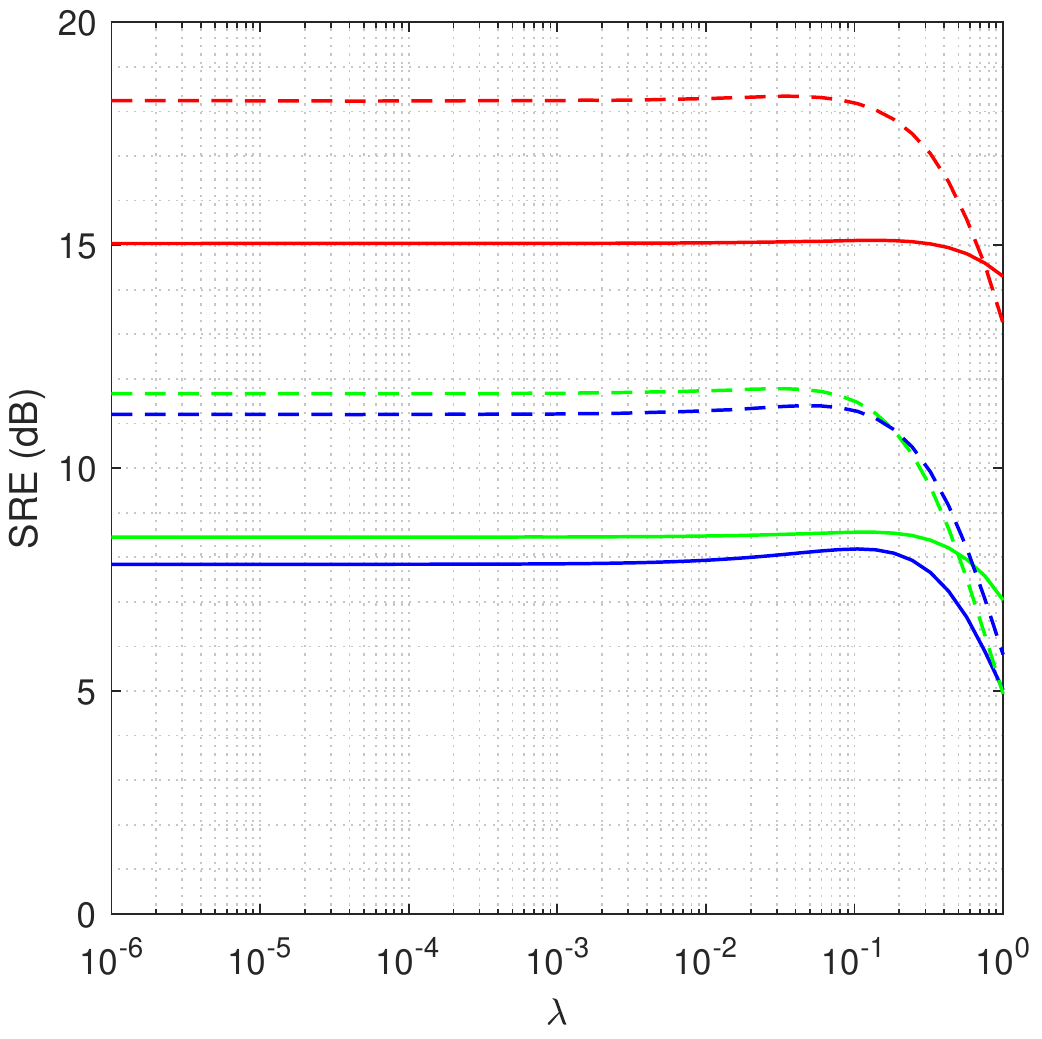}}
		\label{fig:DCs_sensitivity_lambda}
	\end{subfigure}
	\begin{subfigure}[b]{0.29\textwidth}
		\centerline{\includegraphics[width=10em]{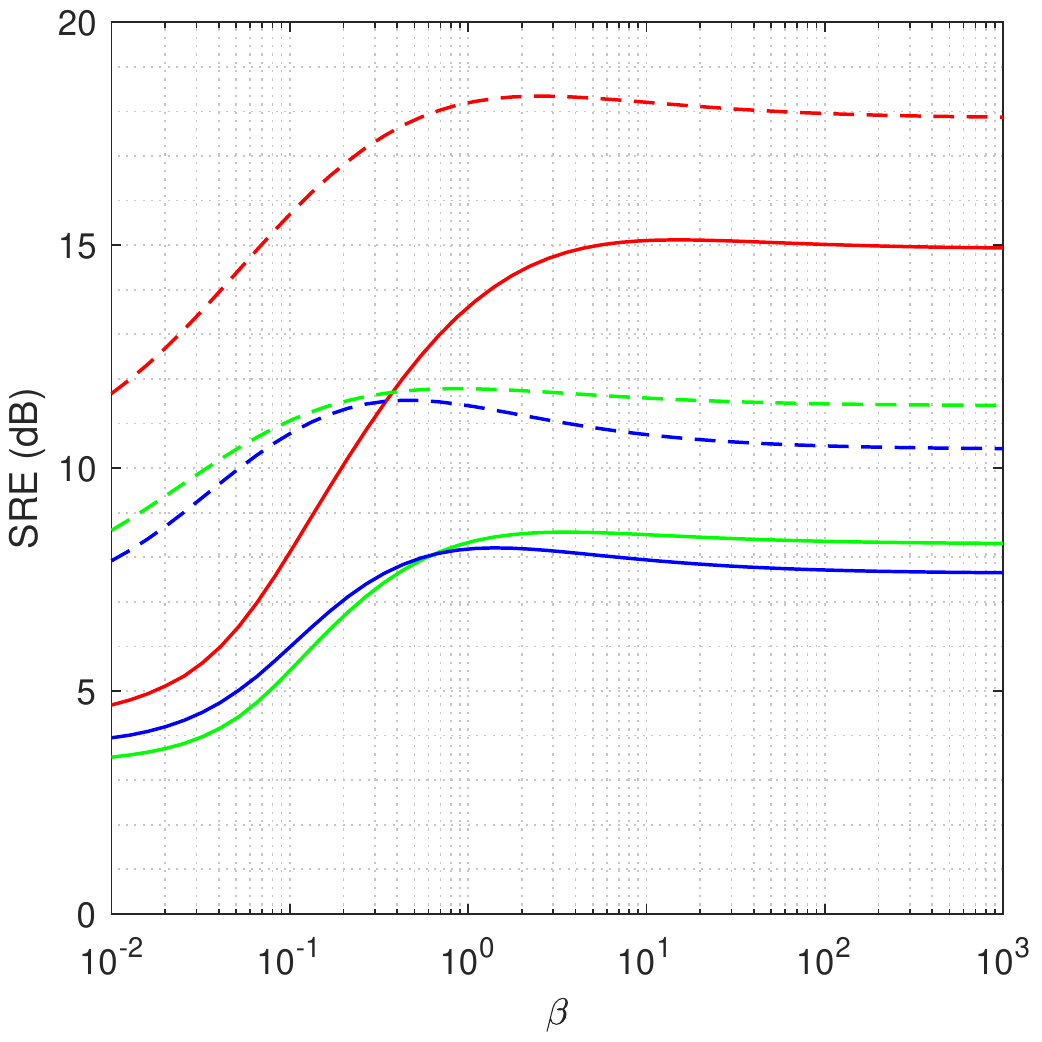}}
		\label{fig:DCs_sensitivity_beta}
	\end{subfigure}
	
	\caption{Variation of the SRE results due to changes in each parameter individually.}
	\label{fig:DCs_sensitivity}
\end{figure}

In this section, the sensitivity of the SRE result to the variation of the HMUA input parameters is analyzed in the unmixing of the DC1, DC2 and DC3 images, for the 20 dB and 30 dB SNR conditions.
With the other parameters set at their optimal values (Table \ref{tab:parameters_HMUA}), the plots in Figure \ref{fig:DCs_sensitivity} show the variation of the method's SRE as a function of a wide range of values for each individual parameter.
The optimal values of $\sigma_1$, $\sigma_2$ and $\sigma_3$ were not changed, as they present an almost uniform progression and are restricted to $\sigma_1 > \sigma_2 > \sigma_3$.
As can be noted, in general, the SRE value remains nearly constant for a consistent range of $\gamma$, $\sigma_0$ and $\lambda$. The most significant reductions occur for $\gamma>0.1$ and $\lambda>0.3$. These are respectively caused by the excessive prioritizations of spatial over spectral regularity in the formation of superpixels, and of sparsity over the reconstruction error (between the real and estimated spectrum of the pixels) during sparse SU. Variations in $\sigma_0$ are compensated by the other rounds of oversegmentation with smaller sizes of superpixels. The enhancement in the quality of spectral unmixing by the outliers removal strategy in the homogeneity assessment step can be seen in the increased SRE obtained by using $\tau_{\text{outliers}}=10\%$ as opposed to $\tau_{\text{outliers}}=0\%$, with differences varying between 0.5 dB and 3 dB.

\begin{table}[h!] 
	\small
	\centering
	\caption{Parameters of the HMUA for each HI.}
	{\begin{tabular}{c||c|c||c|c||c|c}
			\hline
			\hline
			\multicolumn{7}{c}{HMUA} \\
			\hline
			\hline
			\multirow{2}{*}{Parameters} & \multicolumn{2}{c||}{DC1} & \multicolumn{2}{c||}{DC2} & \multicolumn{2}{c}{DC3} \\
			\cline{2-7}
			& 30 dB & 20 dB & 30 dB & 20 dB & 30 dB & 20 dB \\
			\hline
			\hline
			$\gamma$ & 0,00425 & 0,00425 & 0,00025 & 0,00025 & 0,00225 & 0,00225 \\
			\hline
			$\sigma_0$ & 8 & 12 & 6 & 7 & 7 & 8 \\
			\hline
			$\sigma_1$ & 7 & 6 & 5 & 6 & 6 & 7 \\
			\hline
			$\sigma_2$ & 3 & 3 & 4 & 4 & 4 & 4 \\
			\hline
			$\sigma_3$ & 2 & 2 & 2 & 2 & 2 & 3 \\
			\hline
			$\tau_{\text{ outliers}}$ & 10\% & 10\% & 10\% & 10\% & 10\% & 10\%\\
			\hline
			$\tau_{\text{ homog}}$ & 50\% & 20\% & 20\% & 20\% & 30\% & 20\%\\
			\hline
			$\lambda_{\mathcal{C}}$ & 0,003 & 0,007 & 0,003 & 0,007 & 0,005 & 0,01\\
			\hline
			$\lambda$ & 0,03 & 0,1 & 0,03 & 0,1 & 0,05 & 0,1 \\
			\hline
			$\beta$ & 3 & 10 & 3 & 3 & 1 & 1 \\
			\hline
			\hline
	\end{tabular}}
	\label{tab:parameters_HMUA}
\end{table}

\begin{figure}[h!]
	\centering
	\begin{subfigure}[b]{0.31\textwidth}
		\centerline{\includegraphics[width=12em]{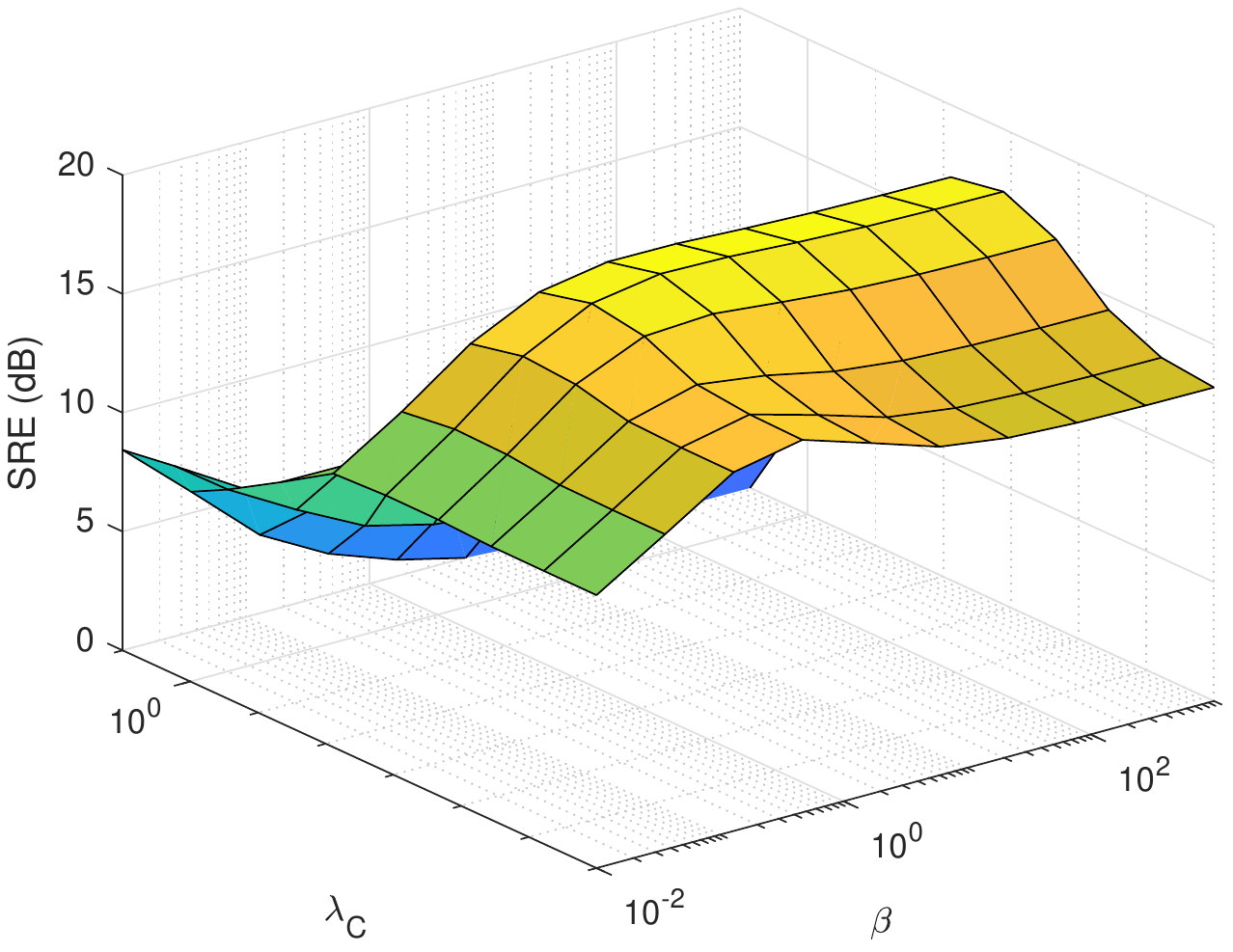}}
		\label{fig:IMG1_30dB_LAMBDA1_BETA}
	\end{subfigure}
	\begin{subfigure}[b]{0.31\textwidth}
		\centerline{\includegraphics[width=12em]{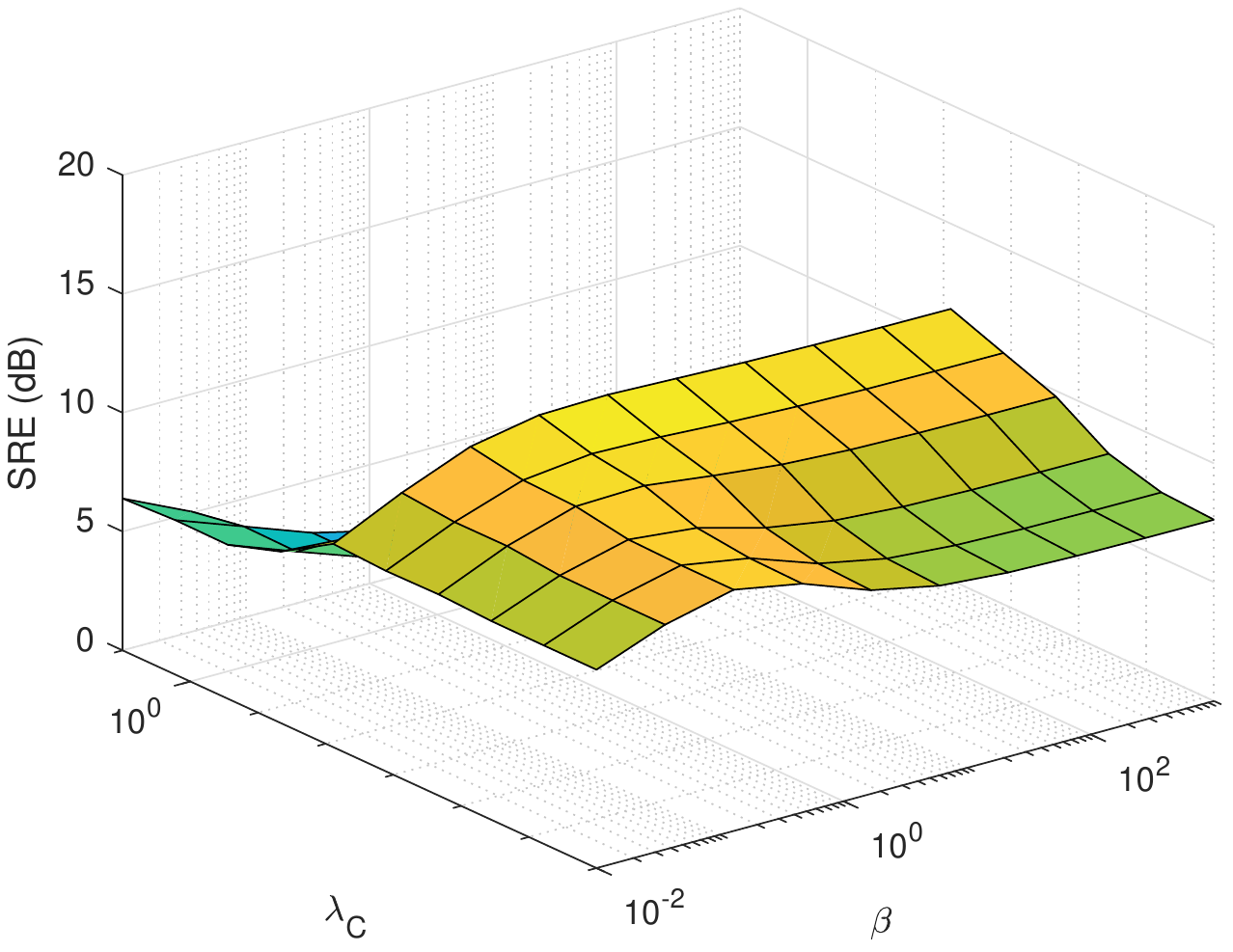}}
		\label{fig:IMG2_30dB_LAMBDA1_BETA}
	\end{subfigure}
	\begin{subfigure}[b]{0.31\textwidth}
		\centerline{\includegraphics[width=12em]{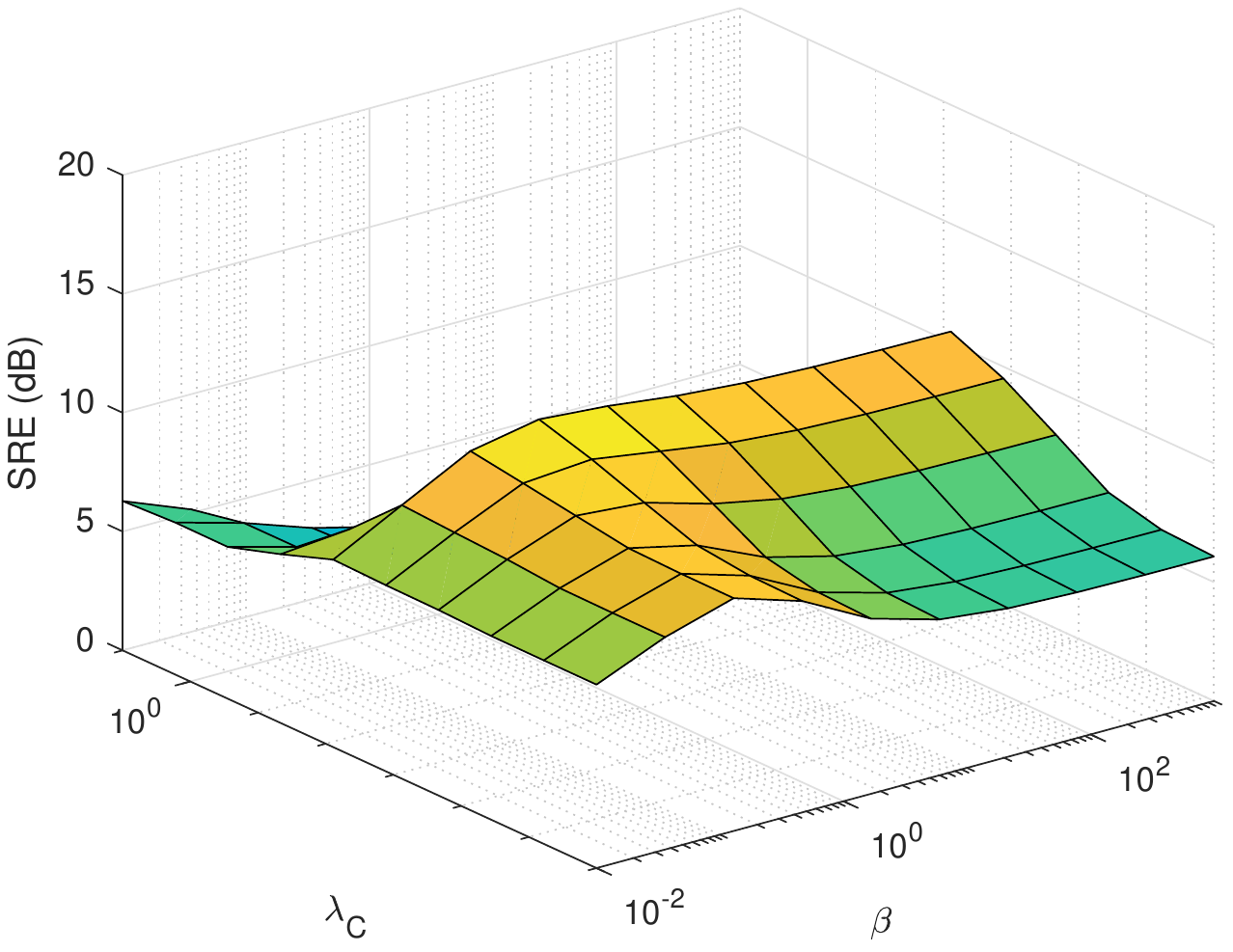}}
		\label{fig:IMG3_30dB_LAMBDA1_BETA}
	\end{subfigure}
	
	\begin{subfigure}[b]{0.31\textwidth}
		\centerline{\includegraphics[width=12em]{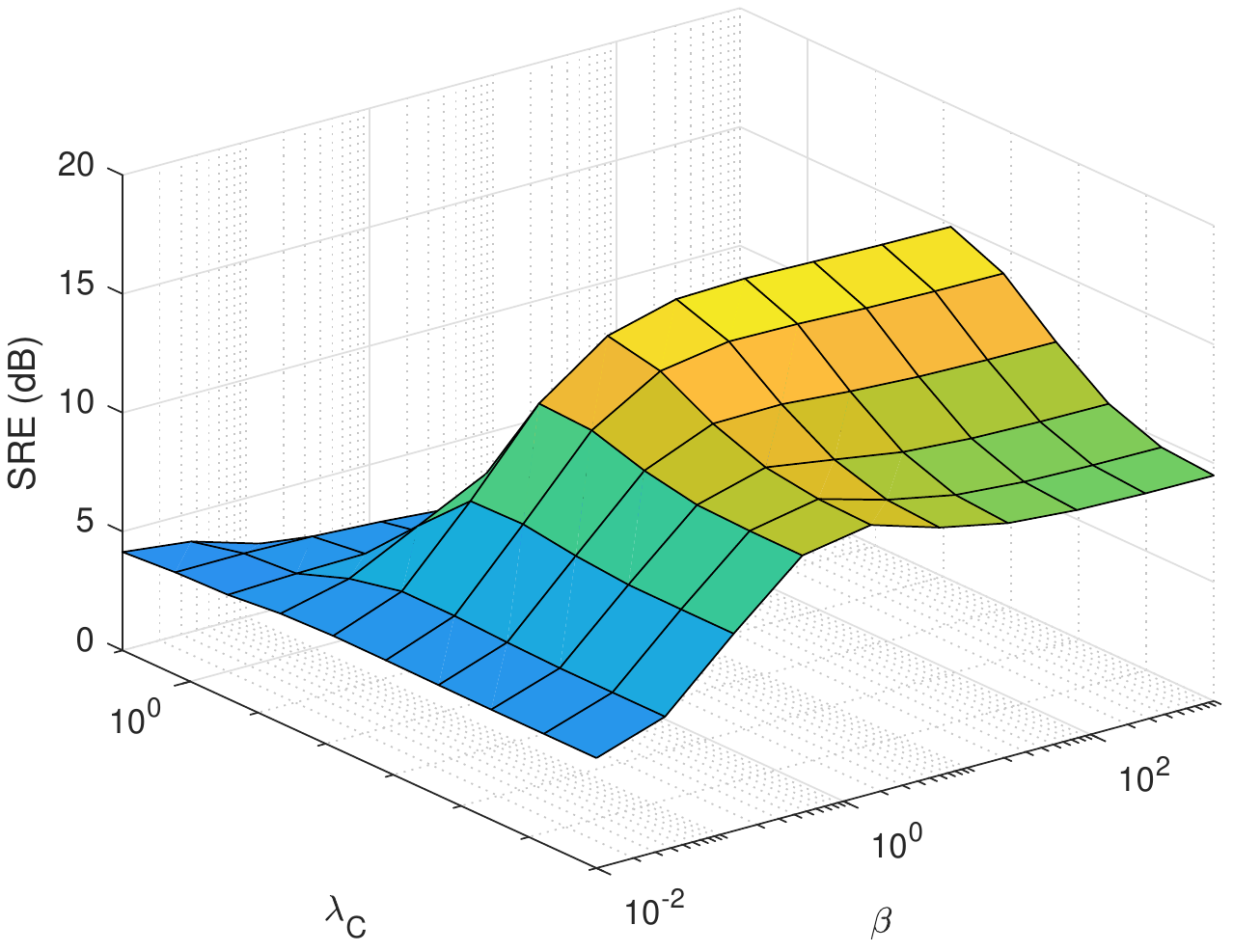}}
		\caption{DC1}
		\label{fig:IMG1_20db_LAMBDA1_BETA}
	\end{subfigure}
	\begin{subfigure}[b]{0.31\textwidth}
		\centerline{\includegraphics[width=12em]{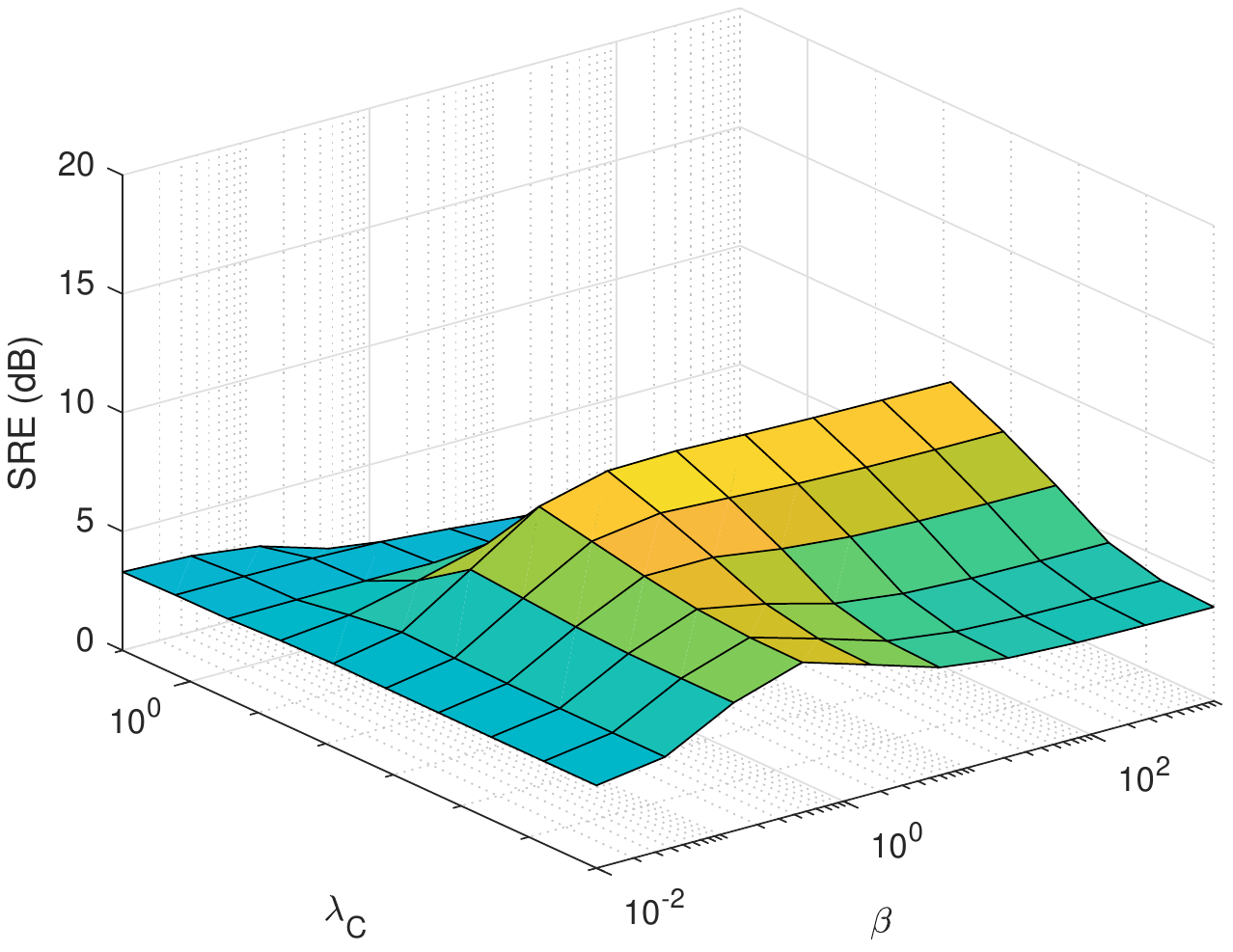}}
		\caption{DC2}
		\label{fig:IMG2_20db_LAMBDA1_BETA}
	\end{subfigure}
	\begin{subfigure}[b]{0.31\textwidth}
		\centerline{\includegraphics[width=12em]{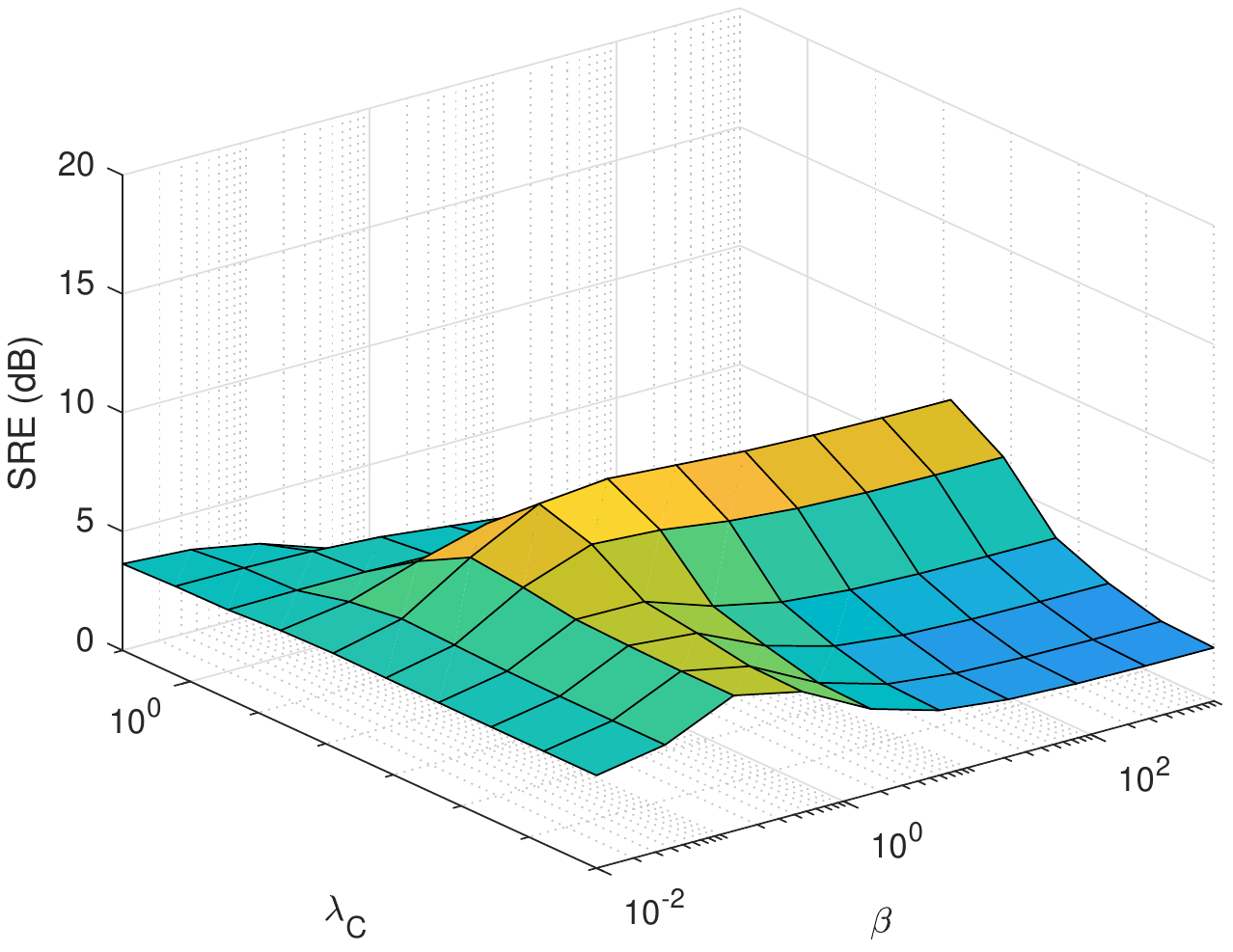}}
		\caption{DC3}
		\label{fig:IMG3_20db_LAMBDA1_BETA}
	\end{subfigure}
	\caption{Joint variation for the optimization parameters $\lambda_{C}$ and $\beta$ for SNR 20 dB (bottom row) and 30 dB (top row) in DC3.}
	\label{fig:IMGs_sens_opt_LAMBDA1_BETA}
\end{figure}

The parameters $\lambda_C$ and $\beta$ show greater influence on the result of the spectral unmixing. In this regard, the joint variation of the parameters $\lambda_C$ $\times$ $\beta$ is shown in Figure \ref{fig:IMGs_sens_opt_LAMBDA1_BETA}. The impact of the variation of the values of $\lambda_C$ and $\beta$ in the SRE show the relevance of the first estimate of abundances obtained in the coarse scale of the image.

\section{Statistical performance analysis} 

As the HMUA has several parameters to be set, we evaluated the chance of an acceptable initial choice of values, without a grid search, based on an appropriate range of response observed empirically for the tested images. HMUA was executed 500 times for each image. Through a uniform distribution, in all executions a new random value for each parameter was chosen according to the following intervals: regularizer $\gamma \in [0,001,\ 0,02]$, superpixels size $\sigma_0 \in [5,\ 20]$, e thresholds $\tau_{outliers} \in [10\%,\ 20\%]$ and $\tau_{homog} \in [10\%,\ 50\%]$ -- regularizers $\lambda_C \in [0,001,\ 0,009]$, $\lambda \in [0,01,\ 0,9]$, $\beta \in [1,\ 50]$. For $\sigma_1$, $\sigma_2$ and $\sigma_3$ the following relationship was used: $\sigma_i \in [\frac{\sigma_{i-1}}{2},\ \sigma_{i-1}-1]$, rounded up to the upper integer and $\sigma_2$ and $\sigma_3$ limited to 3 and 2, respectively. 
After spectral unmixing, the deviation between the SRE value obtained in each execution and the optimal SRE value in Table 2 of [A] was calculated as:

\begin{equation}
	\frac{\text{SRE}-\text{SRE}_{\text{optimal}}}{\text{SRE}_{\text{optimal}}} \times 100\% \,.
\end{equation}

\begin{figure}[h!]
	\centering
	\begin{subfigure}[b]{0.31\textwidth}
		\centerline{\includegraphics[width=11em]{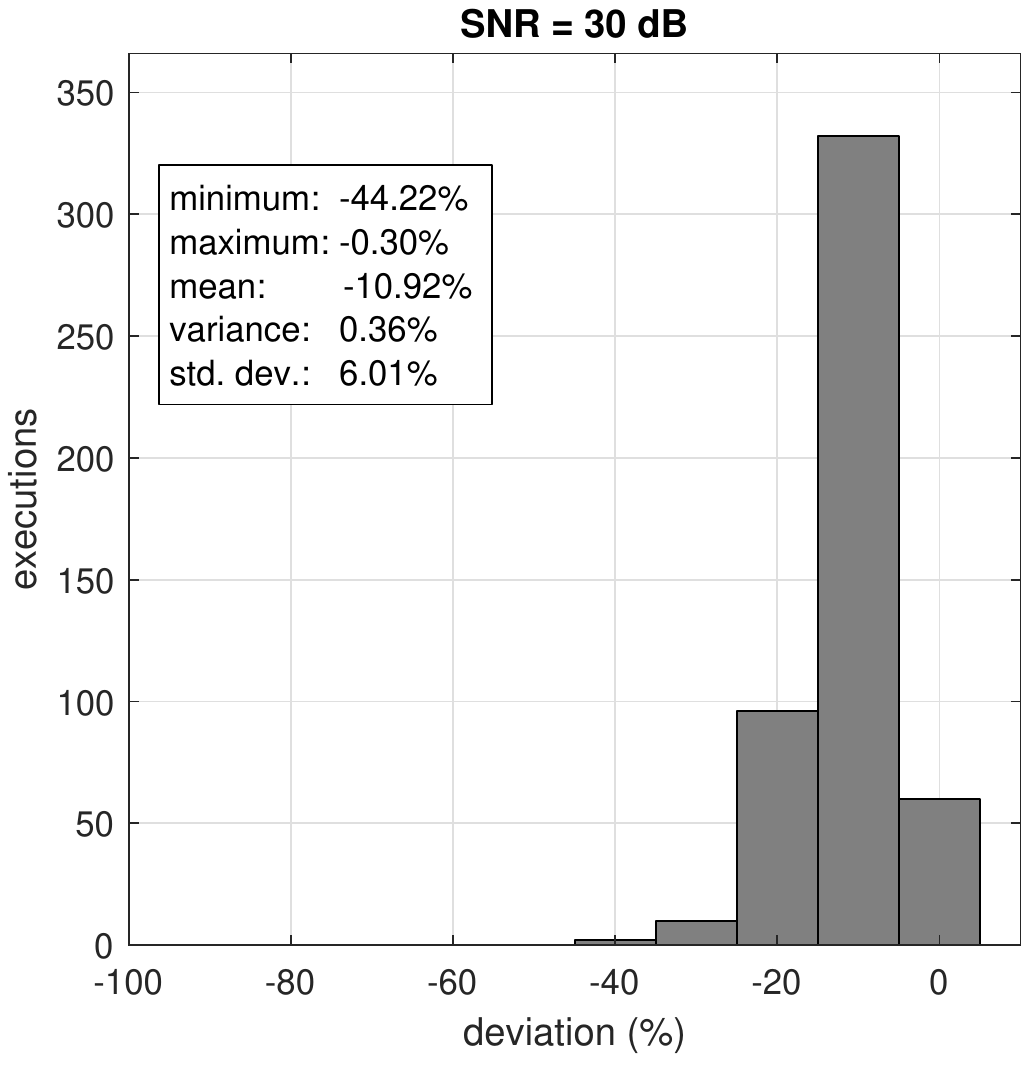}}
		\label{fig:IMG1_30dB_statistical_hist}
	\end{subfigure}
	\begin{subfigure}[b]{0.31\textwidth}
		\centerline{\includegraphics[width=11em]{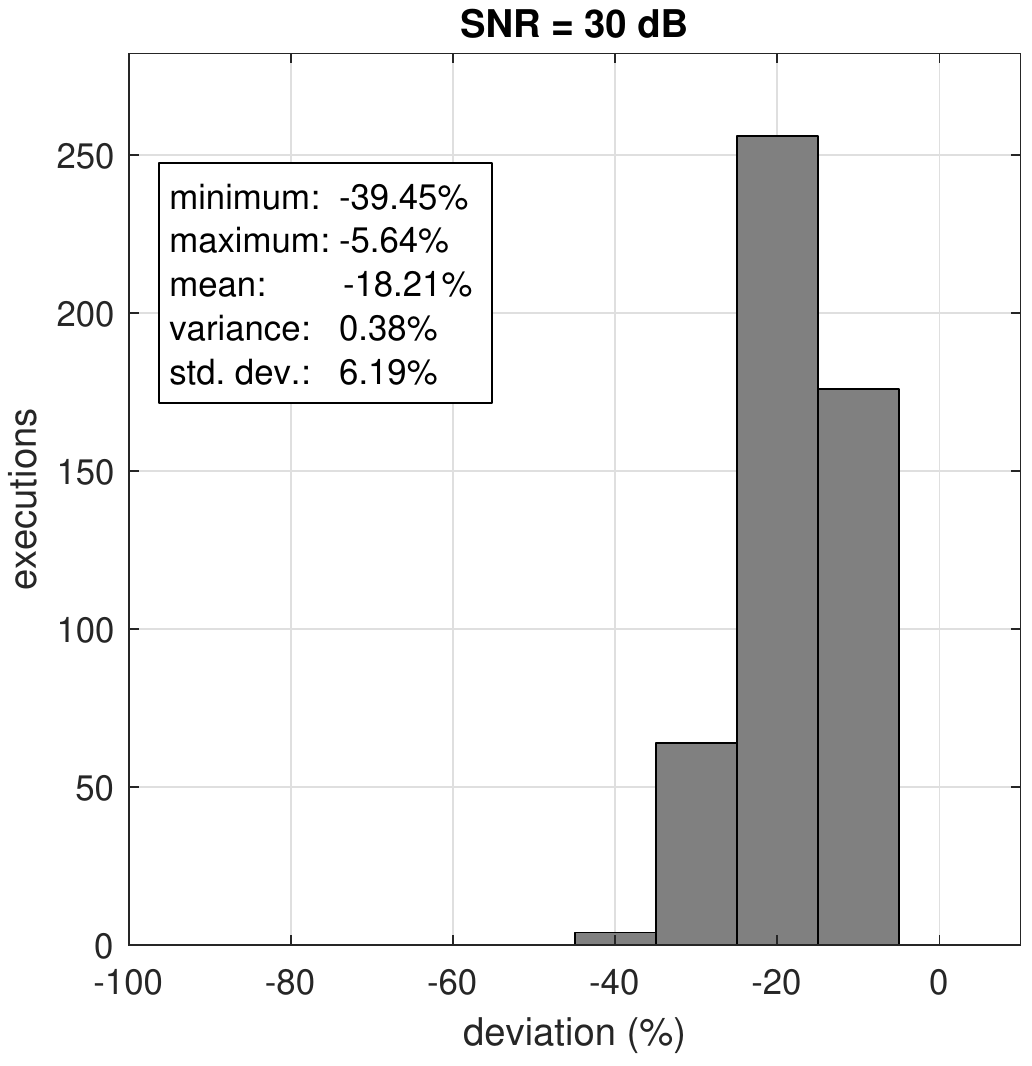}}
		\label{fig:IMG2_30dB_statistical_hist}
	\end{subfigure}
	\begin{subfigure}[b]{0.31\textwidth}
		\centerline{\includegraphics[width=11em]{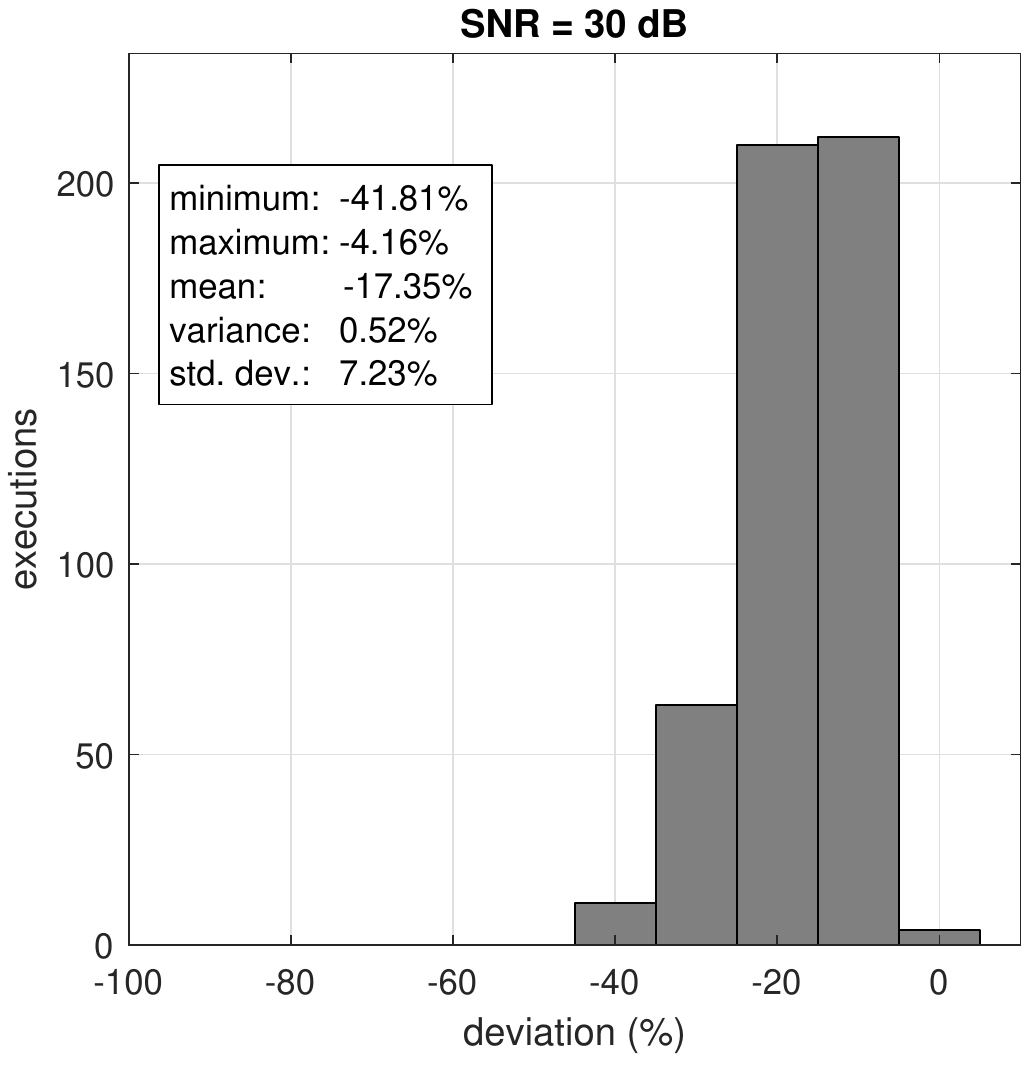}}
		\label{fig:IMG3_30dB_statistical_hist}
	\end{subfigure}
	
	\begin{subfigure}[b]{0.31\textwidth}
		\centerline{\includegraphics[width=11em]{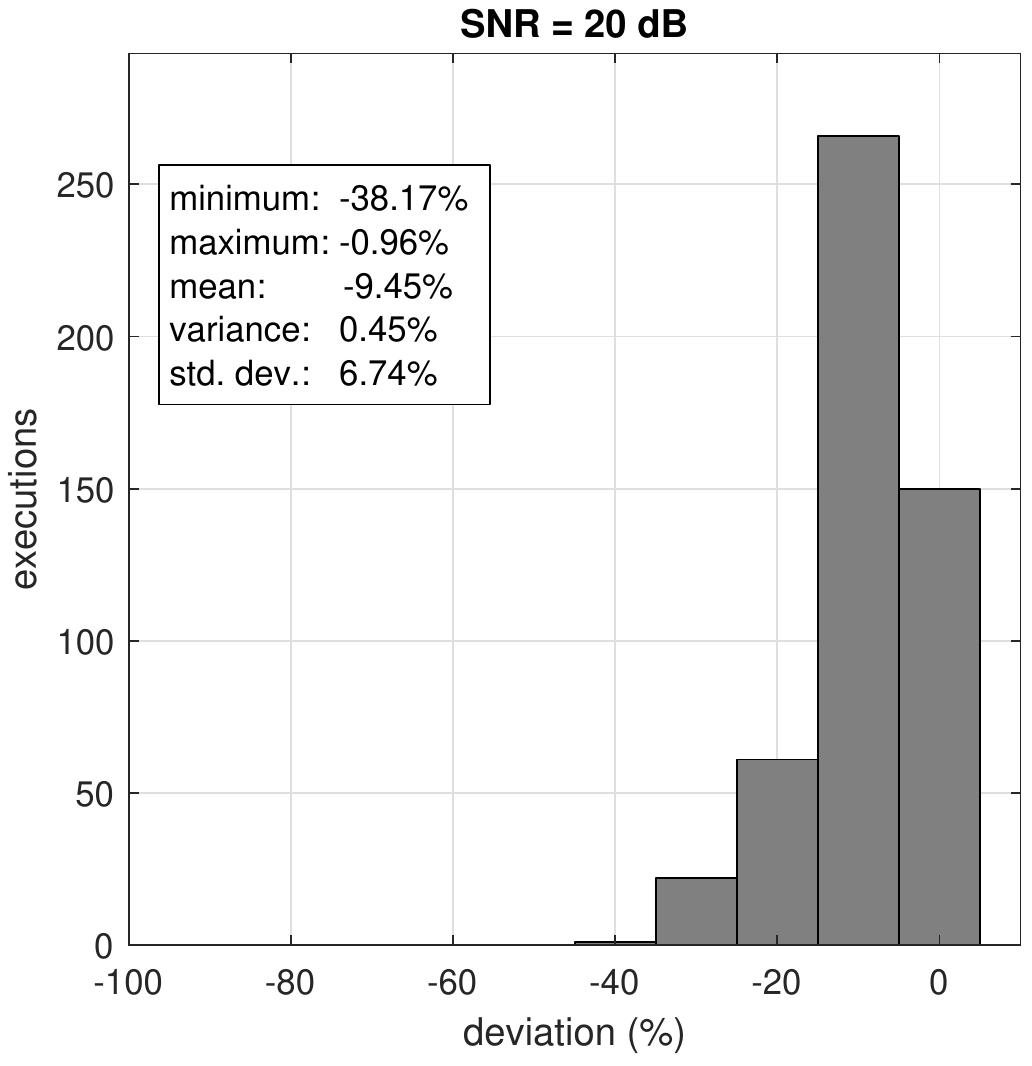}}
		\caption{DC1}
		\label{fig:IMG1_20dB_statistical_hist}
	\end{subfigure}
	\begin{subfigure}[b]{0.31\textwidth}
		\centerline{\includegraphics[width=11em]{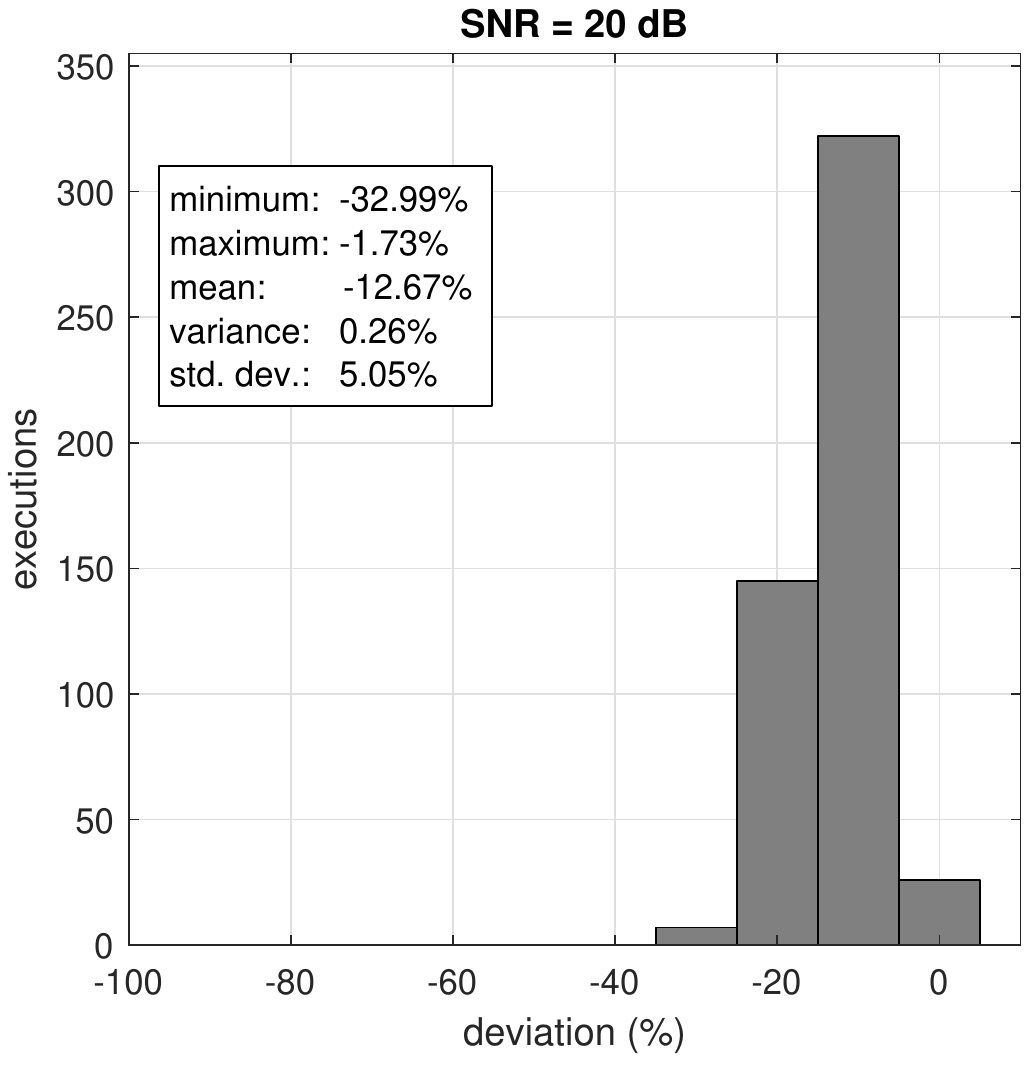}}
		\caption{DC2}
		\label{fig:IMG2_20dB_statistical_hist}
	\end{subfigure}
	\begin{subfigure}[b]{0.31\textwidth}
		\centerline{\includegraphics[width=11em]{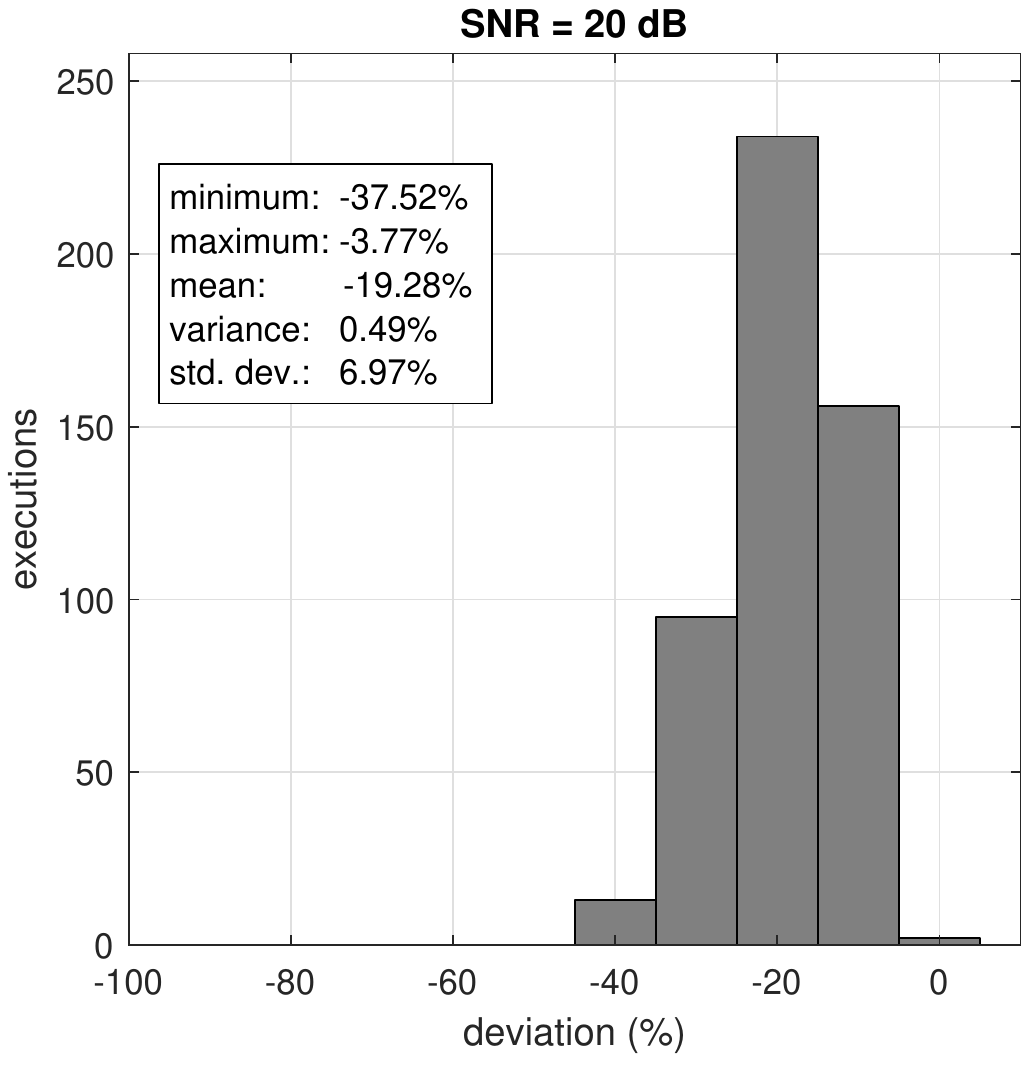}}
		\caption{DC3}
		\label{fig:IMG3_20dB_statistical_hist}
	\end{subfigure}
	\caption{Deviation from the optimal value of the proposed method for randomly chosen parameter values.}
	\label{fig:IMGs_statistical_hist}
\end{figure}

Figure \ref{fig:IMGs_statistical_hist} shows the result of the variations. Through histograms, it is possible to perceive an average variation between 10\% and 20\% and a low standard deviation, around 6\%. This shows that the probability of obtaining good spectral unmixing results is high for parameter values chosen within a reasonable range.

\section{Experiments with real hyperspectral data}

Now, we use well-known real hyperspectral\footnote{Available online at http://lesun.weebly.com/hyperspectral-data-set.html} data, namely, the Samson and Jasper Ridge HIs, to verify the performance of the HMUA method in spectral unmixing, as well as to compare it with the $\text{MUA}_\text{SLIC}$ and S$^2$WSU algorithms. 
To reduce the computation times of the simulations and to make the evaluation of the results easier, two subimages were extracted from these scenes.
Samson's subimage corresponds to a region of 40 $\times$ 95 pixels, with 156 bands between the wavelengths of 401 and 889 nm, composed of three endmembers: Soil, Tree and Water. For Jasper Ridge the chosen region of interest has 50 $\times$ 50 pixels, with 198 bands between 380 and 2500 nm, presenting four main signatures: Road, Soil, Water and Tree. The hyperspectral libraries employed in this work are the same used in [18] of [A], constructed from a technique that extracts $\mathbf{A}$ straight from the HI, resulting in $\mathbf{A} \in \mathbb{R}^{156 \times 105}$ for the Samson and $\mathbf{A} \in \mathbb{R}^{198 \times 529}$ for the Jasper Ridge subimages. For HMUA, $R=1$ was chosen, the other parameters used in each algorithm, shown in Table \ref{tab:parameters_real}, were selected based on the visual comparison between their estimated abundance maps and the RGB representation of the subimages.  

\begin{table}[h!] 
	\small
	\centering
	\caption{Parameters of the algorithms for the real HIs.}
	{\begin{tabular}{c||c||c}
			\hline
			\hline
			Parameters & Samson & Jasper Ridge \\
			\hline
			\hline
			\multicolumn{3}{c}{HMUA} \\
			\hline
			\hline
			$\gamma$ & 0,00125 & 0,00125 \\
			\hline
			$\sigma_0$ & 15 & 15 \\
			\hline
			$\sigma_1$ & 7 & 8 \\
			\hline
			$\tau_{\text{ outliers}}$ & 10\% & 10\% \\
			\hline
			$\tau_{\text{ homog}}$ & 120\% & 100\% \\
			\hline
			$\lambda_{\mathcal{C}}$ & 0,1 & 0,003 \\
			\hline
			$\lambda$ & 0,01 & 0,03 \\
			\hline
			$\beta$ & 1 & 3 \\
			\hline
			\hline
			\multicolumn{3}{c}{$\text{MUA}_\text{SLIC}$} \\
			\hline
			\hline
			$\gamma$ & 0,00125 & 0,00125 \\
			\hline
			$\sigma$ & 7 & 7 \\
			\hline
			$\lambda_{\mathcal{C}}$ & 0,1 & 0,003 \\
			\hline
			$\lambda$ & 0,01 & 0,03 \\
			\hline
			$\beta$ & 1 & 3 \\
			\hline
			\hline
			\multicolumn{3}{c}{S$^2$WSU} \\
			\hline
			\hline
			$\lambda_{\text{swsp}}$ & 0,002 & 0,01 \\
			\hline
			\hline
	\end{tabular}}
	\label{tab:parameters_real}
\end{table}

\begin{figure}[h!]
	\centering
	\begin{subfigure}[b]{0.15\textwidth}
		\centerline{\includegraphics[width=6em]{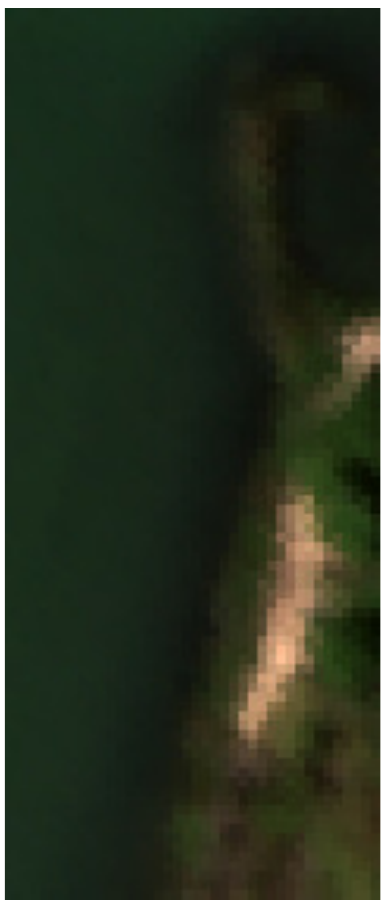}}
		\caption{}
	\end{subfigure}
	\begin{subfigure}[b]{0.15\textwidth}
		\centerline{\includegraphics[width=6em]{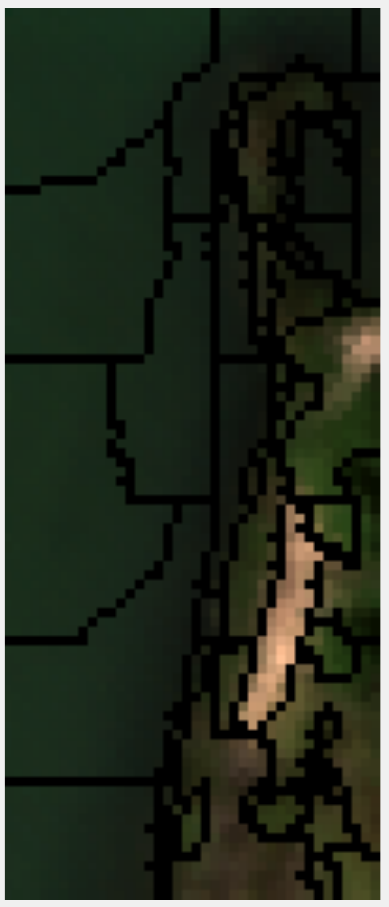}}
		\caption{}
	\end{subfigure}
	\begin{subfigure}[b]{0.15\textwidth}
		\centerline{\includegraphics[width=6em]{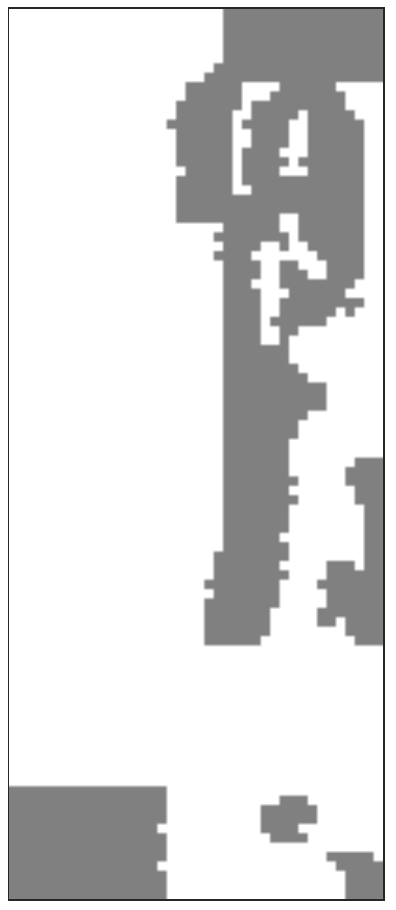}}
		\caption{}
	\end{subfigure}
	\begin{subfigure}[b]{0.15\textwidth}
		\centerline{\includegraphics[width=6em]{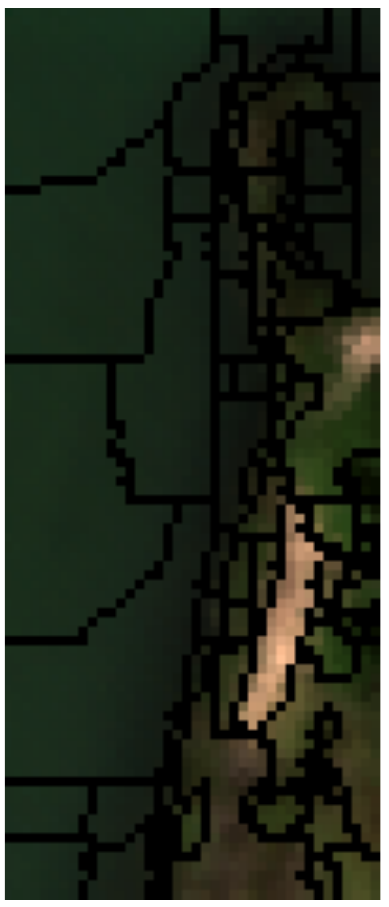}}
		\caption{}
	\end{subfigure}
	
	\caption{(a) Samson subimage RGB representation. (b) Initial oversegmentation, $\sigma_0 = 15$. (c) Initial homogeneity map (non-homogeneous regions in gray color). (d) Result of the final oversegmentation from the subdivision of the non-homogeneous superpixels in (c), $\sigma_1 = 7$.}
	\label{fig:samson_HMUA}
\end{figure}

\begin{figure}[h!]
	\centering
	\begin{subfigure}[b]{0.22\textwidth}
		\centerline{\includegraphics[width=10em]{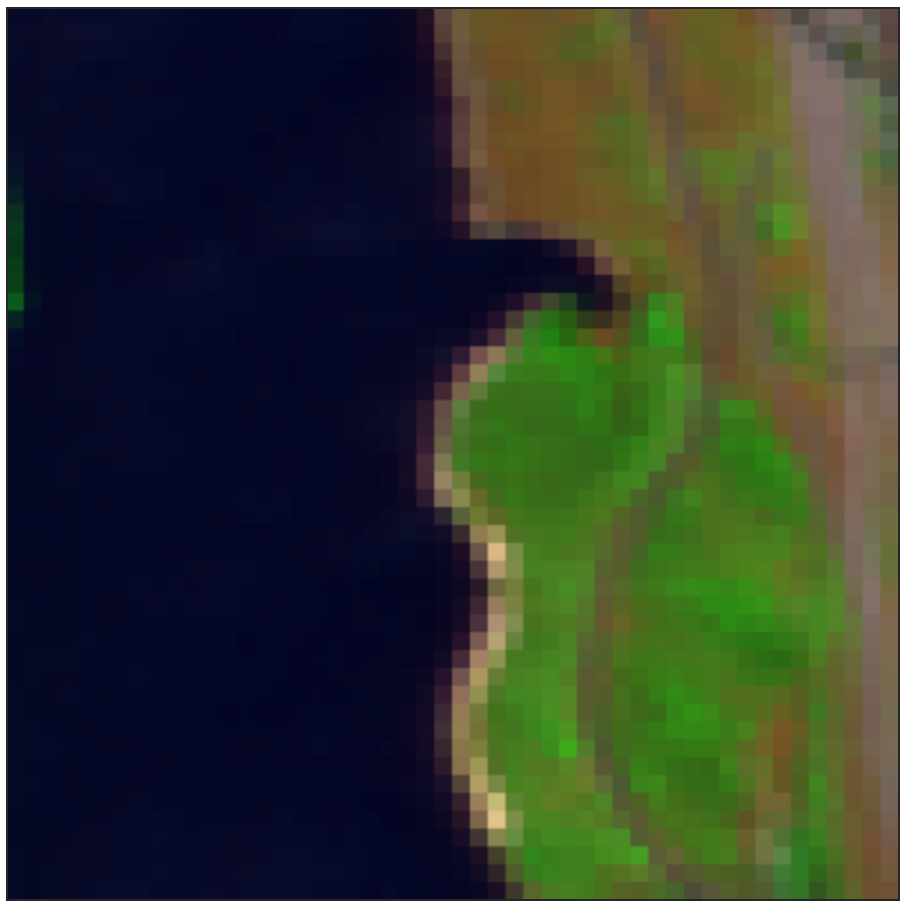}}
		\caption{}
	\end{subfigure}
	\begin{subfigure}[b]{0.22\textwidth}
		\centerline{\includegraphics[width=10em]{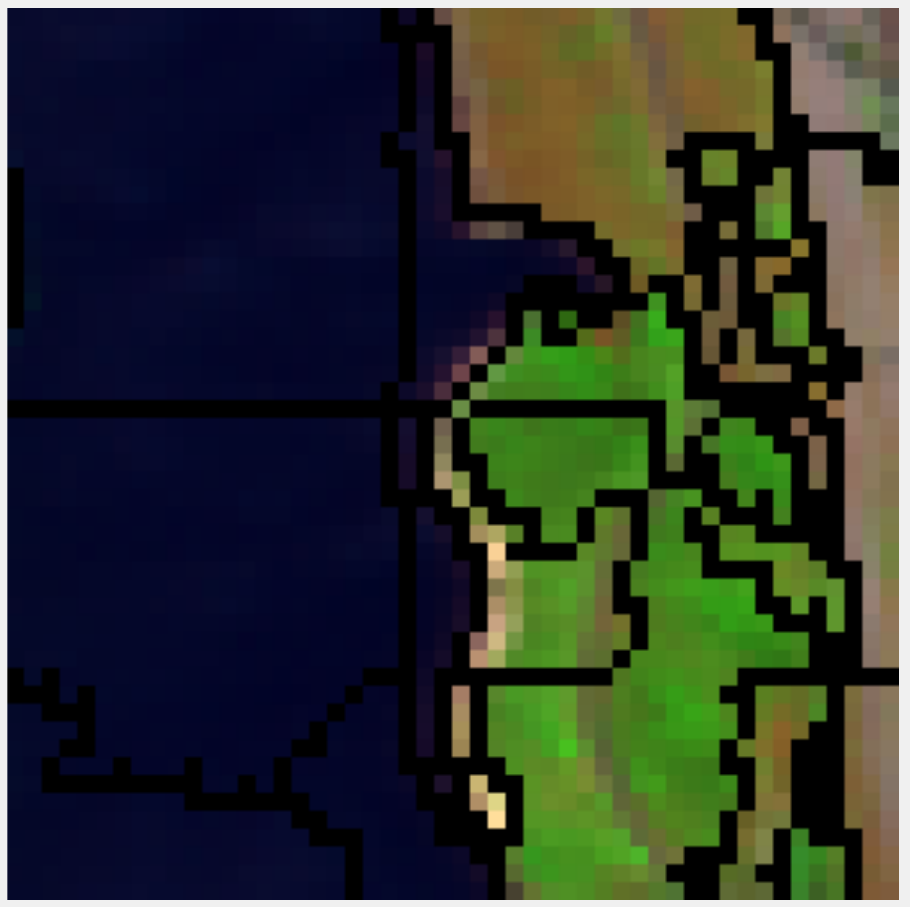}}
		\caption{}
	\end{subfigure}
	\begin{subfigure}[b]{0.22\textwidth}
		\centerline{\includegraphics[width=10em]{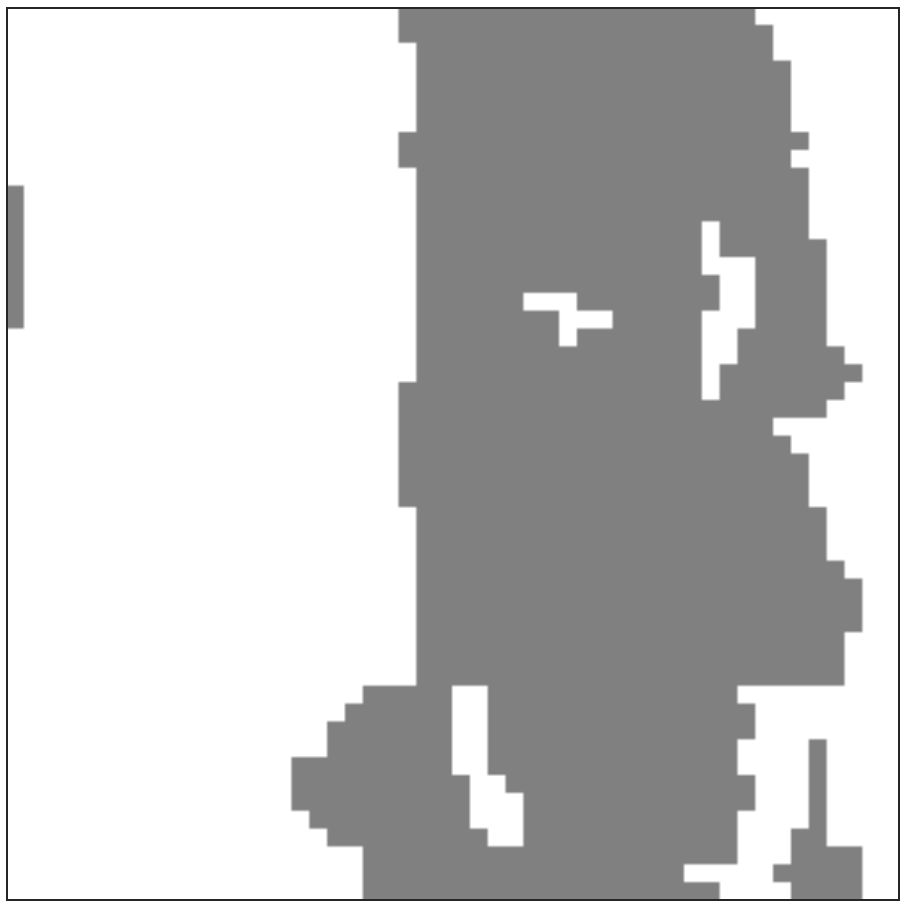}}
		\caption{}
	\end{subfigure}
	\begin{subfigure}[b]{0.22\textwidth}
		\centerline{\includegraphics[width=10em]{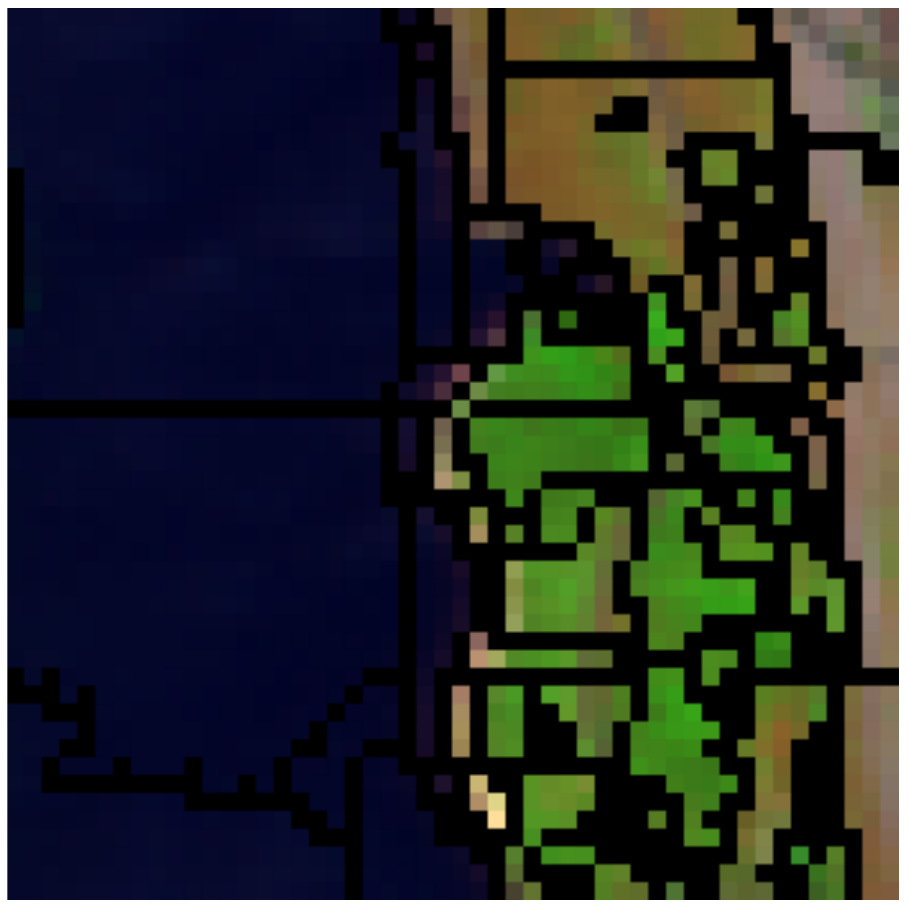}}
		\caption{}
	\end{subfigure}
	
	\caption{(a) Jasper Ridge subimage RGB representation. (b) Initial oversegmentation, $\sigma_0 = 15$. (c) Initial homogeneity map (non-homogeneous regions in gray color). (d) Result of the final oversegmentation from the subdivision of the non-homogeneous superpixels in (c), $\sigma_1 = 8$.}
	\label{fig:jasper_HMUA}
\end{figure}

Figures \ref{fig:samson_HMUA} and \ref{fig:jasper_HMUA} show the oversegmentation results of the Samson and Jasper Ridge subimages, respectively. The effectiveness of the proposed method becomes evident when we observe large superpixels in more uniform regions, as in water, and smaller superpixels in more irregular regions. In these conditions, as with the synthetic image DC1, the strategy proposed in HMUA allows the representation of the image with a significantly smaller amount of superpixels, if compared with $\text{MUA}_\text{SLIC}$, as can be seen in Table \ref{tab:comparare_real}. Also for the simulations with real data, the execution time remains close to the $\text{MUA}_\text{SLIC}$ and substantially faster than the S$^2$WSU.

\begin{table}[h!]
	\centering
	\caption{Average execution time (in seconds) and number of superpixels for the real HIs.}
	\centering
	{\begin{tabular}{c||c||c|c||c|c} 
			\hline
			\hline
			\multirow{2}{*}{Data} & S$^2$WSU & \multicolumn{2}{c||}{$\text{MUA}_\text{SLIC}$} & \multicolumn{2}{c}{HMUA} \\
			\cline{2-6}
			& ex. time & ex. time & superpixels & ex. time & superpixels \\
			\hline
			\hline
			Samson & 9 s & \textbf{5} s & 84 & 7 s & \textbf{51}  \\
			\hline
			\hline
			Jasper Ridge & 19 s & \textbf{7} s & 64 & 9 s & \textbf{42} \\
			\hline
			\hline
	\end{tabular}}
	\label{tab:comparare_real}
\end{table}

\begin{figure}[h!]
	\centerline{\includegraphics[width=30em]{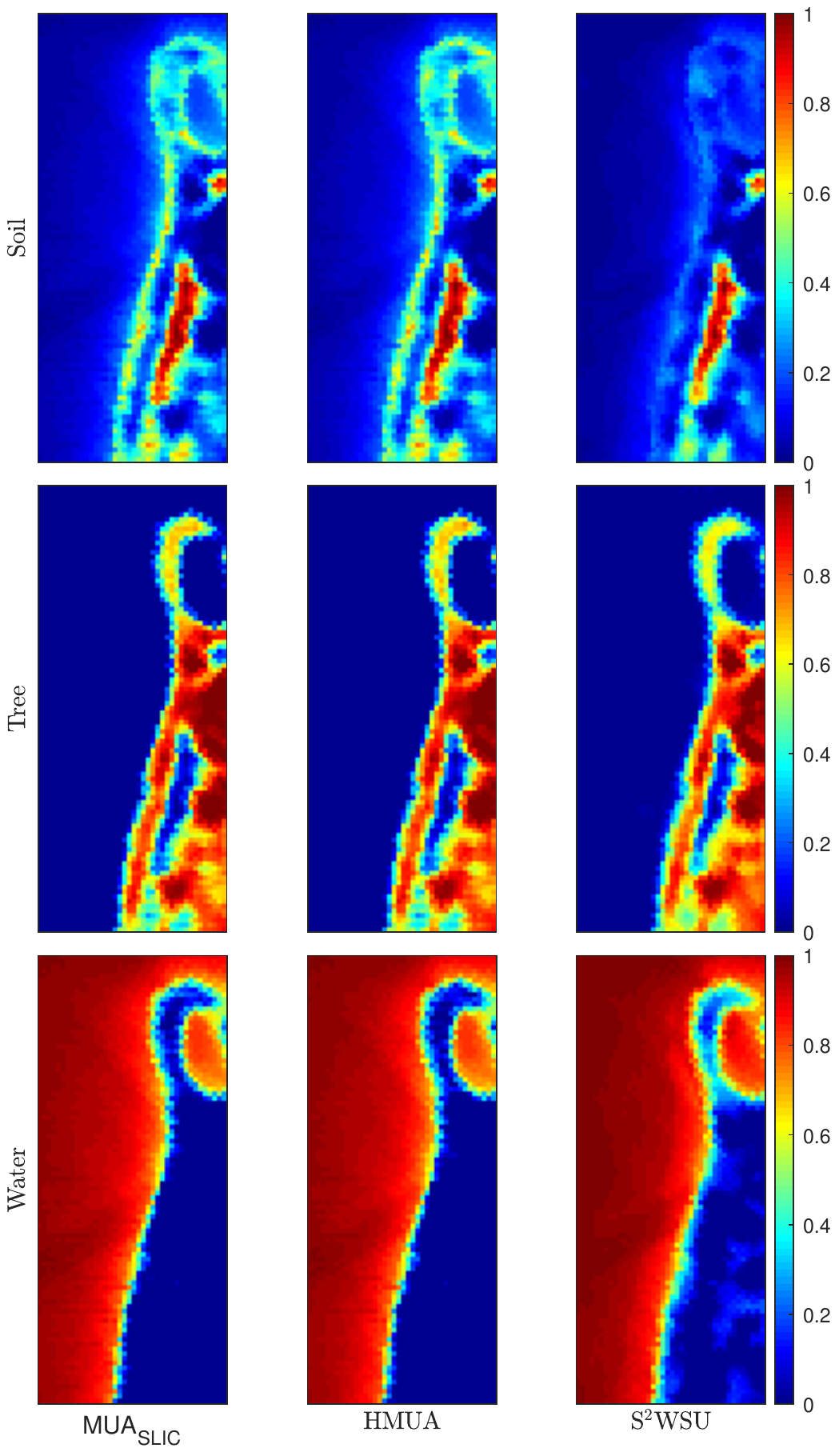}}
	\caption{Samson -- comparison between the estimated abundance maps for each endmember and algorithm.}
	\label{fig:samson_compare}
\end{figure}

\begin{figure}[h!]
	\centerline{\includegraphics[width=35em]{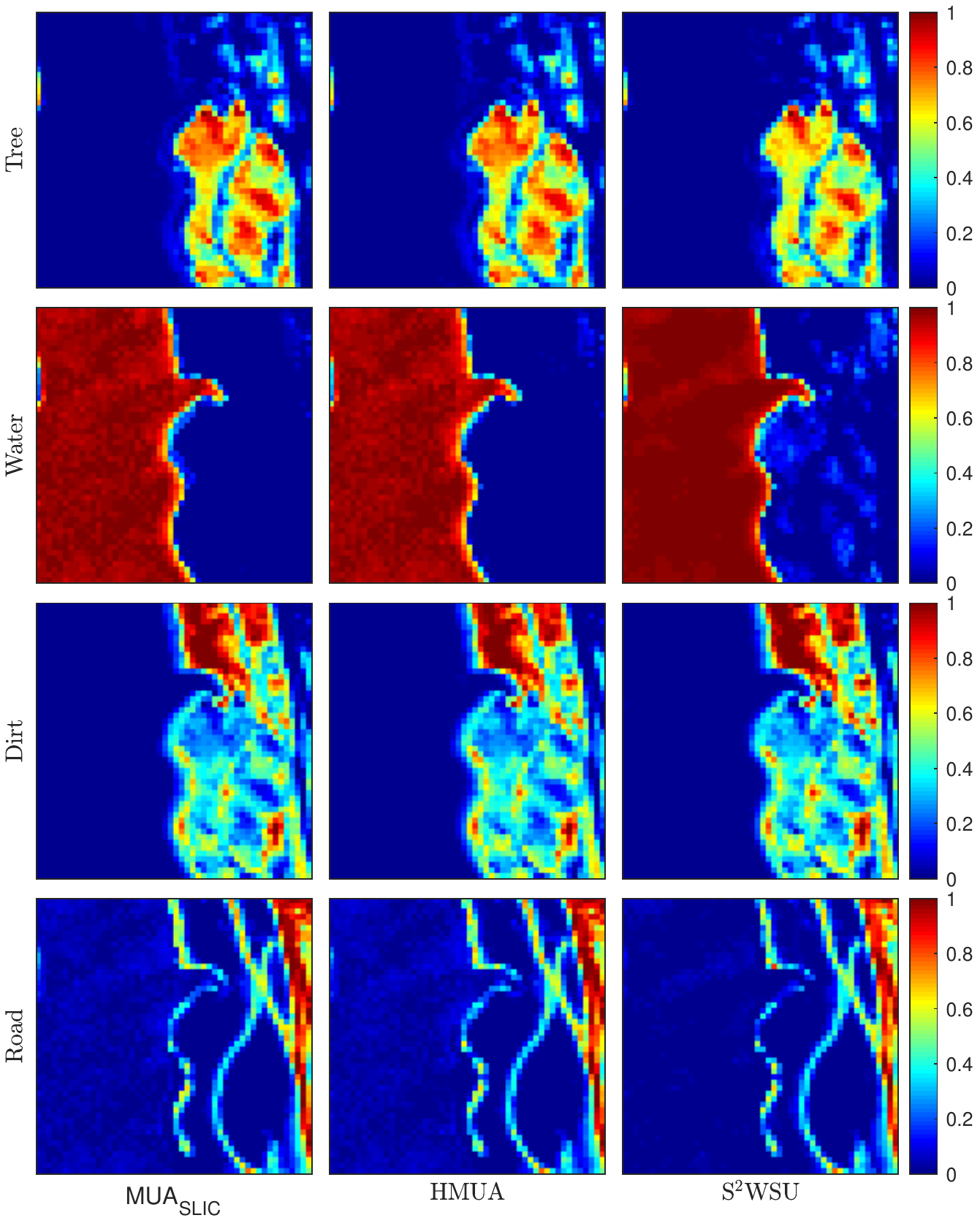}}
	\caption{Jasper Ridge -- comparison between the estimated abundance maps for each endmember and algorithm.}
	\label{fig:jasper_compare}
\end{figure}

It is worth mentioning that the quantitative analysis of spectral unmixing techniques based on experiments with real hyperspectral data is limited, since ground-truth abundance maps are not available for the popular data sets shared on the internet.
Thus, the evaluation of abundance estimation results with real data is restricted to visual inspection of the abundance maps. Therefore, we can notice through the abundance maps estimated in Figures \ref{fig:samson_compare} and \ref{fig:jasper_compare}, that this simpler representation does not compromise the quality of the spectral unmixing. Although the results are visually very similar among the algorithms, if we could quantify the SRE, it is possible that the proposed method would present a slightly better result than the $\text{MUA}_\text{SLIC}$, as occurred in the simulation with all others synthetic data.

\end{document}